\documentclass[12pt]{iopart}
\usepackage{iopams}

 \expandafter\let\csname equation*\endcsname\relax
 \expandafter\let\csname endequation*\endcsname\relax

\usepackage{amsmath}
\usepackage{amssymb}

\usepackage{graphicx}
\usepackage{color}

\definecolor{blue}{rgb}{0,0,1}
\def    \bse{\begin{subequations}}
\def    \ese{\end{subequations}}
\def \be{\begin{equation}}
\def \ee{\end{equation}}
\def \bew{\begin{widetext}\begin{equation}}
\def \eew{\end{equation}\end{widetext}}
\def \bmlett{\begin{mathletters}}
\def \emlett{\end{mathletters}}

\def \ra{\rightarrow}

\def \hrho{\hat{\rho}}

\begin{document}

%\title{State transfer in opto-electro-mechanical systems}
\title{Using dark modes for high-fidelity optomechanical quantum state transfer}
\author{Ying-Dan Wang and Aashish~A.~Clerk}
\address{Department of Physics, McGill University, 3600 rue University, Montreal, QC
Canada H3A 2T8}
%\author{Aashish~A.~Clerk}
%\address{Department of Physics, McGill University, 3600 rue University, Montreal, QC
%Canada H3A 2T8}
%\begin{abstract}
%In a recent publication [Y.D. Wang and A.A. Clerk, Phys. Rev. Lett. \textbf{108}, 153603 (2012)],
%we have shown the existence of a mechanically-dark
%mode can significantly increase the state transfer fidelity between a
%microwave cavity and an optical cavity mediated by a commonly-coupled
%mechanical resonator. In this article, we elaborate on this issue and provide technical details of the double swap and
%adiabatic passage schemes. Moreover, we proposed a new transfer scheme: hybrid
%scheme, which utilizes both the dark mode (immune to the mechanical noise) and the
%bright mode (exposed to the mechanical noise). The hybrid scheme combines
%the advantage of fast operation and robust to mechanical thermal noise.
%Particularly, this new scheme enables high-fidelity state transfer even
%without pre-cooling the mechanics into ground state. We analyzed in detail
%the transfer fidelity and the features of each scheme and provide a ``phase
%diagram'' to the optimal transfer scheme in different experimental
%conditions. We also provide technique derivations for the transfer of
%itinerant state, both Gaussian and non-Gaussian. Especially, we give more
%detailed explanation of the EIT physics in this system.
%\end{abstract}
\begin{abstract}
In a recent publication [Y.D. Wang and A.A. Clerk, Phys.~Rev.~Lett.~\textbf{108}, 153603 (2012)],
we demonstrated that one can use interference to significantly increase the
fidelity of state transfer between two electromagnetic cavities coupled to a common mechanical resonator
over a naive sequential-transfer scheme based on two swap operations.  This involved making use of a delocalized electromagnetic
mode which is decoupled from the mechanical resonator, a so-called ``mechanically-dark" mode.
Here, we demonstrate the existence of a
new ``hybrid" state transfer scheme which incorporates the best elements of the dark-mode scheme (protection
against mechanical dissipation) and the double-swap scheme (fast operation time). Importantly, this new scheme also does not require
the mechanical resonator to be prepared initially in its ground state.  We also provide additional details on the previously-described
interference-enhanced transfer schemes, and provide an enhanced discussion of how the interference physics here is intimately related to
the optomechanical analogue of electromagnetically-induced transparency (EIT).  We also compare the various transfer schemes over a wide
range of relevant experimental parameters, producing a ``phase diagram" showing the the optimal transfer scheme for different points in parameter space.
\end{abstract}

\date{today}
\maketitle

%\tableofcontents

\section{Introduction}

\label{section:introduction}

The fields of optomechanics and electromechanics are interested in the physics of systems where a
mechanical resonator is coupled to either a driven optical cavity, or a driven electronic circuit. The interaction of an optical cavity and mechanical oscillation due to radiation pressure or photon thermal effect has been studied in early experiments (see. e.g.~\cite{Metzger2004,Carmon2005}). In the past few years, these fields have achieved significant progress--
ground state cooling and many-photon strong coupling have been realized
in experiments on both optomechanical and electromechanical systems~\cite{Groeblacher2009,Chan2011,Verhagen2012,Teufel2011a,Teufel2011b}. As mechanical resonators can achieve strong coupling with both microwave and optical cavities, this recent experimental progress may soon enable a powerful new application of these systems: the ability to transfer quantum states or traveling photons between optical cavities and microwave-frequency electrical systems. Such optomechanically-enabled state transfer could allow one to combine the complimentary advantages of both microwaves and optics. For example, one could imagine using a superconducting qubit to prepare a highly non-classical state of a microwave cavity (as recently demonstrated in Ref.~\cite{Hofheinz2009}), and then transfer this state to an optical cavity, which provides an interface for long-distance information propagation via photons.
Optomechanical state transfer could also serve as a key component in quantum information processing networks which combine a variety of different elements over long distances (see, e.g.,~\cite{Stannigel2010}).

In such schemes, one would be interested in the ability to transfer intra-cavity states, as well as the states of itinerant photons incident on one cavity. To be explicit, by intra-cavity state transfer we mean that the state to be transferred is the initial ($t=0$) state of the first cavity; this could even be a non-classical state prepared by, e.g., having cavity 1 interacting with a qubit at $t<0$. The goal is now to design a protocol so that cavity 2 ends by being in this same state. In contrast, for itinerant photon transfer schemes, the goal is to have a wave packet incident on cavity 1 be faithfully reproduced at the output of cavity 2. The bandwidth of this wave packet could be much much smaller than the damping rates of the cavities.  Such conversion of itinerant photons of completely different wave length has various intriguing applications and has been discussed in several recent works~\cite{Safavi2011njp,Wang2012,Tian2012}. In this paper, we will analyze both of the two transfer tasks.

The problem of transferring quantum states between a mechanical resonator and cavity photons was the subject of several early studies~\cite{Parkins1999,Zhang2003}, utilizing strategies well-known in the field of cavity QED~\cite{Zeng1994}. While the systems considered there involve only two resonators (an optical cavity and a mechanical resonator), similar
approaches could be used in a three-resonator system where, e.g., both a microwave and an optical cavity are jointly coupled to a mechanical resonator; this is the subject of more recent studies~\cite{Tian2010,Regal2011,Safavi2011njp}.
For the problem of transferring intra-cavity states, these works consider the use of
two successive ``swap" operations in a two-cavity optomechanical system (Fig.~\ref{fig:schematics}). One pulses the optomechanical interactions such that initially, only the first cavity and the mechanical resonator are coupled.  The coupling
is then left on for just long enough to exchange the states of the first cavity and the mechanical resonator. This is then repeated to exchange the
mechanical and the second cavity states (i.e. the cavity 1 - mechanical resonator coupling is turned off, the cavity 2 - mechanical resonator coupling is turned on)
\cite{Tian2010,Regal2011}. Note that single-swap operations between an optical cavity and mechanical resonator have recently been demonstrated experimentally~\cite{Verhagen2012,Fiore2011}. While straightforward, the ``double-swap" scheme for intra-cavity state transfer has the drawback that as the state to be transferred occupies the mechanical resonator during the transfer protocol, it will be strongly degraded by any thermal noise driving the mechanical resonator.  This remains true even if one first prepares the mechanical resonator in the ground state.

In a recent work~\cite{Wang2012}, we addressed the above problem by now looking at the two-cavity optomechanical system as a true three-mode system, as opposed to only considering two modes (i.e.~one cavity mode and the mechanical resonance) at a time. One finds that there exists in general a delocalized cavity eigenmode of the coherent optomechanical Hamiltonian which is decoupled from the mechanical resonator, the so-called ``mechanically-dark" mode. Using this mode enables high
fidelity state transfer even in the presence of strong mechanical dissipation; similar ideas were also studied in~\cite{Tian2012}.
In this paper, we provide additional technical details of the transfer protocols discussed in Ref.~\cite{Wang2012}, in particular
details on the analytic calculation of transfer fidelities, as well as a thorough discussion of the dependence of the transfer fidelity on the
initial state of the mechanical resonator.  Further, we introduce a new scheme to deal with one of the key difficulties of the adiabatic transfer scheme discussed in Ref.~\cite{Wang2012}, namely that to adiabatically use the dark mode for state transfer, the transfer protocol must be very slow, and hence is extremely susceptible to degradation from cavity losses.  The new scheme we describe is a so-called ``hybrid" scheme, as it only partially involves the dark-mode described above.  While this means that it is not completely immune to mechanical heating effects, we show that it allows for a much faster transfer time, and is hence less susceptible to cavity losses. The hybrid scheme also has the surprising advantage that it does not require the mechanical resonator to be initially prepared in its ground state; we provide a heuristic explanation for this insensitivity in the text. As we analyze in detail in the paper,
the hybrid scheme works best for typical experimental parameters where both cavity dissipation and mechanical noise are equally relevant.
We give a comparison to the different features of each intra cavity state transfer scheme. We also discuss the transfer of
itinerant photons and the underlying optomechanical EIT mechanism.

The remainder of this paper is organized as follows.  The basic system Hamiltonian and the definition of the
dark mode are discussed in Sec.~2. In Sec.~3, we introduce and describe three basic schemes to implement intra-cavity state transfer:
the double-swap scheme of Ref.~\cite{Tian2010} (which does not use the dark mode),
the adiabatic passage scheme of Refs.~\cite{Wang2012,Tian2012} (which entirely uses the dark mode), and the new hybrid scheme
(which only partially uses the dark mode).  The fidelity and features of each scheme is analyzed. We also compare the three
schemes and provide a ``phase diagram'' showing the optimal transfer scheme for various choices of experimental parameters (optomechanical coupling
dissipation, temperature, etc.).  In Sec.~4, we change focus to now consider the state transfer of itinerant photons, that is wave-packets incident
on one of the two cavities; this expands upon the analysis given in Ref.~\cite{Wang2012}. Particular attention is given to clarifying the connection between
the high fidelities possible here and optomechanical analogue of electromagnetically-induced transparency (EIT)~\cite{Agarwal2010,Weis2010,Safavi2011}.
Conclusions and remarks are provided in Sec.~5.

\section{Basics of the two-cavity optomechanical setup}

\label{section:hamiltonian}

\subsection{Setup and Hamiltonian}

\label{subsec:hamiltonian}

\begin{figure}[tbp]
\begin{center}
\includegraphics[width= 0.8 \columnwidth]{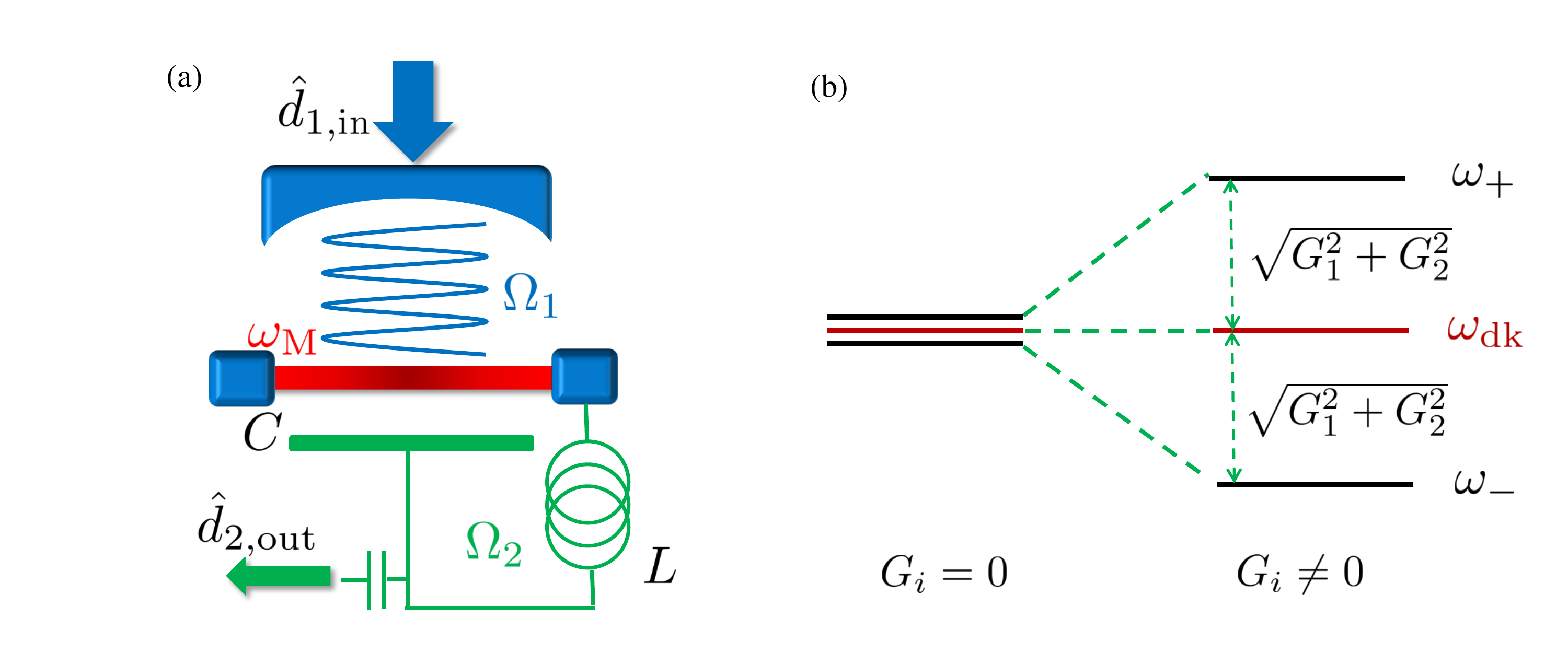}
\end{center}
\caption{(Color online) (a) The two-cavity optomechanical system where a
mechanical resonator (red bar in the middle) is coupled simultaneously to an
optical cavity (the Fabry-Perot cavity in blue) and a microwave cavity (the LC
circuit in green). Possible experimental realizations of this setup has been described in detail in Ref.~\cite{Regal2011,Safavi2011njp}. (b) The energy diagram of the three eigenmodes of the system for $G_i=0$ (left) and $G_i\neq 0$ (right). $G_i$ is the driven coupling strength of the mechanical resonator to the two cavities which is tunable with the drives (see Eq.~(\ref{ham}) and the following text for more details). The middle red line corresponds to the mechanically-dark mode, whose frequency is independent of the coupling. $\sqrt{G_1^2+G_2^2}$ is the energy difference between the dark mode and the two mixed modes.}
\label{fig:schematics}
\end{figure}

For concreteness, we consider an optomechanical system where a single mechanical resonator is
simultaneously coupled to both an optical cavity and a microwave cavity via
dispersive couplings (see Fig.~\ref{fig:schematics}(a)); particular
experimental realizations are discussed in Ref.~\cite{Regal2011,Safavi2011njp}. The cavities interact with the mechanical
resonator through the standard radiation pressure coupling ($\hbar = 1$)
\begin{equation}
\hat{H}_{{\rm int},i}=g_{i}\hat{b}_{i}^{\dag }\hat{b}_{i}\left( \hat{a}+\hat{a}^{\dag
}\right) ,
\end{equation}%
where $\hat{b}_{i}$ is the annihilation operator of cavity $i$ ($i=1,2$), $\hat{a%
}$ is the annihilation operator of the mechanical resonator, $g_{i}$ is the
single-photon optomechanical coupling strength to cavity $i$, and the mechanical frequency is $%
\omega _{M}$. Note that to date, almost all experimentally studied devices are firmly in the limit of a weak
single-photon optomechanical coupling, $g_i \ll \kappa_i$, the exception being experiments where the collective motion of cold-atoms
in a cavity acts as the effective mechanical degree of freedom~\cite{Gupta2007}.

We further specialize to the standard situation where each cavity is
strongly driven, resulting in a large average photon number which serves to enhance the
effective cavity-mechanical resonator coupling.  It is useful to separate the cavity state into a classical
amplitude $\tilde{b}_i$ plus a deviation $\hat{d}_i$ which accounts for the effects of noise and the interaction with the mechanics:
\begin{equation}
\hat{b}_{i}=e^{-i\omega _{d,i}t}\left( \tilde{b} _{i}+\hat{d}_{i}\right) ,
\end{equation}%
where $\omega _{d,i}$ is the drive frequency of cavity $i$.
The classical amplitude $\tilde{b} _{i}$ is simply proportional to the amplitude of the laser drive applied to cavity $i$; without loss of generality,
we take it to be real and positive. $\hat{d}_{i}$ can be interpreted as the cavity $i$ annihilation operator in a frame displaced by the classical amplitude.
For a large drive $\tilde{b}_i \gg 1$ and small single-photon optomechanical couplings, we can safely neglect terms in the displaced frame which are quadratic in the displaced cavity operators, resulting in purely linear coupling between each cavity and the mechanical resonator (see, e.g.,~\cite{Marquardt2007,Wilson2007}).  Finally, we work in an interaction picture with respect to the two cavity drives.  The resulting Hamiltonian in this displaced
interaction picture (in the absence of any dissipation) reads:
\begin{equation}
\hat{H}_{0}=\omega _{M}\hat{a}^{\dag }\hat{a}-\sum_{i=1,2}\left[ \Delta _{i}%
\hat{d}_{i}^{\dag }\hat{d}_{i}-G_{i}\left( \hat{a}^{\dag }\hat{d}_{i}+\hat{d}%
_{i}^{\dag }\hat{a}\right) \right]  \label{ham}
\end{equation}%
Here, $\Delta _{i} = \omega_{d,i} - \Omega_{i}$ is the detuning of the drive applied to cavity $i$ and $\Omega_i$ is the cavity resonance frequency; we have assumed detunings near the red-detuned mechanical sideband of each cavity
(i.e.~$\Delta_i \sim -\omega_M$) and anticipated taking the good cavity limit $\kappa_i \ll \omega _{M}$ ($\kappa_i$ is the energy damping rate of cavity $i$), which allows us to make a rotating-wave approximation in writing the optomechanical interaction.
The driven (or many-photon) optomechanical coupling between the mechanical resonator and cavity $i$ is denoted as $G_{i}=\tilde{b} _{i}g_{i}$ (we let $G_{i}>0$ throughout the paper); note that $\tilde{b} _{i}$ is proportional to the drive amplitude applied to cavity $i$,
and thus can be controlled in time.

We see quite clearly from Eq.~(\ref{ham}) that in our displaced frame, the optomechanical Hamiltonian has exactly the form we need for state transfer:  the optomechanical interaction can move quanta from each cavity to the mechanical resonator, and vice-versa.  We will thus be interested in quantum state transfer {\it in the displaced frame} of each cavity. Throughout this paper, the quantum state to be transferred is represented by $\hat{d}_{i}$.  This state of interest thus sits atop a large classical coherent state $| \tilde{b}_i \rangle$ which is used to generate a large effective optomechanical coupling.  Equivalently put, if we want to transfer a given state $|\psi\rangle$ from cavity 1 to cavity 2, cavity 1 should be initially prepared in a state $| \psi' \rangle$ which is just the state $| \psi \rangle$ displaced an amount $\tilde{b}_i$ in phase space.  The role of this phase-space displacement is solely to achieve a large effective optomechanical coupling. Note that these two parts of the cavity state will be spectrally separated:  the background coherent-state part of the cavity field used to generate a coupling is concentrated at the red-detuned sideband (i.e.~$\Omega_{i} - \omega_M$), whereas the state to be transferred is located roughly within a cavity bandwidth $\kappa$ of the cavity resonance frequency $\Omega_{i}$.

We now turn to including dissipation in our system. The two cavities and mechanical resonator are assumed to be coupled to
independent Ohmic baths that describe both internal loss and the couplings
to the extra-cavity modes used to drive each cavity. For simplicity, we take each cavity to be single-sided (i.e.~each cavity has a single input-output port). The master equation for the reduced density matrix $\hrho \left( t\right) $ describing both cavities and the mechanical
resonator takes the form:
\begin{equation}
\dot{\hrho}\left( t\right) =-i\left[ \hat{H},\hrho \left( t\right) \right] +
\mathcal{L}_{\text{C}}\hrho \left( t\right) +\mathcal{L}_{\text{M}}\hrho
\left( t\right)  \label{master}
\end{equation}%
where the dissipation can be described by Lindblad type dissipator
\begin{equation}
\mathcal{L}_{\text{C}}\hrho \left( t\right) =\sum_{i=1,2}\kappa _{i}\left(
N_{i}+1\right) \mathcal{D}\left( \hat{d}_{i}\right) \hrho \left( t\right)
+\sum_{i=1,2}\kappa _{i}\left( N_{i}\right) \mathcal{D}\left( \hat{d}%
_{i}^{\dag }\right) \hrho \left( t\right) ,  \label{lc}
\end{equation}%
\begin{equation}
\mathcal{L}_{\text{M}}\hrho \left( t\right) =\gamma \left( N_{M}+1\right)
\mathcal{D}\left( \hat{a}\right) \hrho \left( t\right) +\gamma \left(
N_{M}\right) \mathcal{D}\left( \hat{a}^{\dag }\right) \hrho \left( t\right) .
\label{ld}
\end{equation}%
Here $\gamma $ ($\kappa _{i}$) denote the energy decay rate of the
mechanical resonator (cavity $i$), and $N_{M}$ ($N_{i}$) denote the bath
temperature, expressed as the number of thermal quanta, of the bath coupled
to the mechanical resonator (cavity $i$). The super-operator $\mathcal{D}(\hat{A})$ is defined as:
\begin{equation}
\mathcal{D}\left( \hat{A}\right) \hrho \left( t\right) \equiv \hat{A}\hrho
\left( t\right) \hat{A}^{\dag }-\frac{1}{2}\left\{ \hat{A}^{\dag }\hat{A}%
,\hrho \left( t\right) \right\} _{+}.
\end{equation}%
%We also assumed the optimal situation where each cavity is far into the
%resolved-sideband regime $\omega _{M}\gg \kappa _{i}$, and where each cavity
%is driven near the red-detuned mechanical sideband (i.e.~$\Delta _{i}\sim
%-\omega _{M}$). This permits us to make a rotating wave approximation in
%writing the optomechanical interactions, resulting in a \textquotedblleft
%beam-splitter" form which is optimal for state transfer~\cite{Parkins1999}.
Finally, note that we are assuming throughout the paper that there is negligible phase noise in the large driving amplitudes on each cavity
used to generate the optomechanical couplings $G_i$.  The effect of
such noise on coherent transfer was studied in Ref.~\cite{Diosi2008,Rabl2009}; the resulting requirements are similar to those needed for cavity cooling~\cite{Rabl2009}, and are thus within reach of experiment.

%%%%%%%%%%%%%%%%%%%%%%%%%%%%%%%%%%%%%%%%%%%

\subsection{Mechanically-dark mode}

\label{subsec:darkmode}

To understand how interference could be used to enhance mechanically-mediated state transfer in our system, we first note that the mechanical resonator only couples to a particular linear combination of the cavity modes. Defining new canonical mode operators
\begin{equation}
\hat{c}_{\text{\textrm{br}}}(t)\equiv \frac{1}{\sqrt{%
G_{1}^{2}(t)+G_{2}^{2}(t)}}\left( G_{1}(t)\hat{d}_{1}+G_{2}(t)\hat{d}%
_{2}\right)  \label{brm}
\end{equation}%
and
\begin{equation}
\hat{c}_{\mathrm{dk}}(t)\equiv \frac{1}{\sqrt{G_{1}^{2}(t)+G_{2}^{2}(t)}}%
\left( -G_{2}(t)\hat{d}_{1}+G_{1}(t)\hat{d}_{2}\right) ,  \label{dkm}
\end{equation}%
the Hamiltonian Eq.~(\ref{ham}) can be rewritten as
\begin{equation}
\hat{H}_{0} =
    \omega _{M} \left( \hat{a}^{\dag }\hat{a}
        + \hat{c}_{\text{\textrm{br}}}^{\dag }\hat{c}_{\text{\textrm{br}}}
        +  \hat{c}_{\mathrm{dk}}^{\dag}\hat{c}_{\mathrm{dk}} \right)
%\hat{H}_{0}=\omega _{M}\hat{a}^{\dag }\hat{a}-\Delta \hat{c}_{\text{\textrm{%
+\sqrt{G_{1}^{2}+G_{2}^{2}}\left( \hat{a}^{\dag }\hat{c%
}_{\text{\textrm{br}}}+\hat{c}_{\text{\textrm{br}}}^{\dag }\hat{a}\right),
\label{h2}
\end{equation}%
where we have assumed the two cavities are each driven at the red-detuned sideband $\Delta _{i}=\Delta = - \omega_M  $. We see that
the ``mechanically-bright" mode described by $\hat{c}_{\text{\textrm{br}}}$ is coupled with the mechanics, whereas
the ``mechanically-dark" mode described by $\hat{c}_{\mathrm{dk}}$ is decoupled from the mechanics.

The above Hamiltonian is trivially diagonalized as
\begin{equation}
\hat{H}_{0}=\sum_{j = +,-, \text{dk}}\hbar \omega _{j}\hat{c}_{j}^{\dag }\hat{c}_{j},
\end{equation}%
where
\begin{equation}
\hat{c}_{\pm }(t)\equiv \frac{1}{\sqrt{2}}\left( \hat{c}_{\text{br}}\left(
t\right) \pm \hat{a}\right),  \label{cpm}
\end{equation}%
describes hybridized modes formed from the mechanical resonator $\hat{a}$,
and the mechanically-bright cavity mode $\hat{c}_{\text{br}}\left( t\right)$, and the mode eigenfrequencies are given by:
\begin{eqnarray}
    \omega _{\pm }      & = & \omega _{M}\pm \sqrt{G_{1}^{2}+G_{2}^{2}}
        \label{eq:omegapm}\\
    \omega_{\text{dk}} & = & \omega_M
\end{eqnarray}
The energy structure of the three modes is shown in Fig.~\ref{fig:schematics}(b).

As the dark mode $\hat{c}_{\mathrm{dk}}(t)$ of the two cavities does not couple to the mechanical resonator (c.f.~Eq.~(\ref{h2})),
one would expect it to have immunity against the effects of mechanical dissipation (c.f.~Eq.~(\ref{ld})).  This simple fact will provide the foundation
for several state transfer schemes that are robust against mechanical dissipation. However, it is worth noting that this protection is not exact when one considers the case of asymmetric cavity dissipation, that is $\kappa_1 \neq \kappa_2$. One can easily understand this by just looking at the damping effects of the cavity dissipation.  The Heisenberg-Langevin equation of motion for each cavity mode operator will have a damping term of the form:
\begin{equation}
    \frac{d}{dt} \hat{d}_{i} = \left(-i \omega_M - \kappa_i/2 \right) \hat{d}_i  + ...
\end{equation}
Some simple algebra now shows that if we re-write these expressions in the eigenmode basis, then for $\kappa_1 \neq \kappa_2$, the damping terms in the equation of motion will mix the dark cavity mode and the bright cavity mode.  The result is that mechanical noise can now corrupt the dark mode: mechanical noise will heat the coupled bright mode, and this noise will then drive the cavity dark mode via the cavity damping terms.  We can see this more explicitly by writing the super-operator of  cavity dissipation in the eigenmode basis:
\begin{eqnarray}
\mathcal{L}_{\text{C}}\hrho \left( t\right) &=&\frac{\kappa _{1}\left(
N_{1}+1\right) +\kappa _{2}\left( N_{2}+1\right) }{2}\left( \mathcal{D}%
\left( \hat{c}_{\text{\textrm{br}}}\right) \hrho \left( t\right) +\mathcal{D}%
\left( \hat{c}_{\text{\textrm{dk}}}\right) \hrho \left( t\right) \right)
\notag \\
&&+\frac{\kappa _{1}N_{1}+\kappa _{2}N_{2}}{4}\left( \mathcal{D}\left( \hat{c%
}_{\text{\textrm{br}}}^{\dag }\right) \hrho \left( t\right) +\mathcal{D}%
\left( \hat{c}_{\text{\textrm{dk}}}^{\dag }\right) \hrho \left( t\right)
\right)  \notag \\
&&+\frac{\kappa _{1}\left( N_{1}+1\right) -\kappa _{2}\left( N_{2}+1\right)
}{2}\mathcal{F}\left( \hat{c}_{\text{\textrm{br}}},\hat{c}_{\text{\textrm{dk}%
}}^{\dag }\right) \hrho \left( t\right)  \notag \\
&&+\frac{\kappa _{1}N_{1}-\kappa _{2}N_{2}}{2}\mathcal{F}\left( \hat{c}_{%
\text{\textrm{dk}}}^{\dag },\hat{c}_{\text{\textrm{br}}}\right) \hrho \left(
t\right)  \label{dbrdk}
\end{eqnarray}%
with the super-operator is defined as%
\begin{equation}
\mathcal{F}\left( \hat{A},\hat{B}\right) \hrho \left( t\right) \equiv \hat{A}%
\hrho \left( t\right) \hat{B}-\frac{1}{2}\left\{ \hat{B}\hat{A},\hrho \left(
t\right) \right\} _{+}.
\end{equation}%
The last two lines of Eq. (\ref{dbrdk}) represent the mixing of the bright
and dark mode; these terms vanish when both $\kappa _{1}=\kappa _{2}$ and $N_{1}=N_{2}$.
This mixing effect leads to an effective heating of the dark mode which is a factor
$\sim (\kappa_1 - \kappa_2)/G$ smaller than the heating of the mechanically-bright cavity mode.
While the mechanically-dark mode is only completely immune from mechanical noise for completely symmetric dissipation, we will show that schemes utilizing it can still be highly effective even when there are deviations from this perfect symmetry condition.
Note that a definition of dark mode for $%
\kappa _{1}\neq \kappa _{2}$ based on a perturbative treatment of the cavity damping asymmetry was
discussed in~\cite{Tian2012}.
%It was also found that the extra heating due to $\kappa_1 \neq \kappa_2$ is $(\kappa_1 - \kappa_2)/G$ times smaller than the direct heating on the bright mode~\cite{Tian2012}.

%, these terms
%vanishes and the dark mode is disentangled from the bright mode. Thus the dark
%mode is completely insensitive to the mechanical thermal noise. Using this
%mode for state transfer gains protection against mechanical thermal
%dissipation. Otherwise, the cavity damping Eq.~(\ref{lc}), can effectively
%mix the dark mode with the bright mode. A definition of dark mode for $%
%\kappa _{1}\neq \kappa _{2}$ based on perturbative treatment is also
%discussed in~\cite{Tian2012}. However, in most cases, this is only a small
%limitation in the regime that is realistic in experiments. We will analyze
%this effect in the subsequent subsections on each transfer scheme.

\section{Intra-cavity state transfer}

\label{section:intracavity}

We first consider protocols which allow the two-cavity optomechanical system in Fig.~\ref{fig:schematics}
to be used to transfer intra-cavity states.  The initial state of interest is prepared in cavity $1$, and the goal is to have this
state end up in cavity $2$ at the end of the transfer protocol with a maximal fidelity.
We will discuss 3 different schemes to implement such a transfer
process: the double-swap protocol (which does not use the dark mode),
the adiabatic passage protocol (entirely based on the dark mode) and the hybrid swap protocol (which only partially uses the dark mode).
In the absence of any cavity or mechanical dissipation, all these schemes allow transfer with perfect fidelity.  However, in the presence of dissipation, this is not so;
moreover, each scheme has its own relative merits and disadvantages versus the different kinds of system dissipation.  We will discuss this in what follows, with the goal
of providing insight into what the optimal strategy is given a certain set of system parameters.

%One important application of the setup is to transfer intra-cavity state
%which is prepared in one cavity. In this way, the advantage of both optical
%cavity and microwave cavity can be combined. For example, by interacting
%with a superconducting phase qubit, an arbitrary state can be prepared in
%microwave cavity~\cite{Hofheinz2009}. Transferring those states to the
%optical cavity can be an efficient way to generate highly non-classical
%optical states. This also sets up a communication platform between the
%solid-state qubits and the flying qubits: photon.

%In order to transfer intra-cavity
%state, strong coupling condition $\kappa \ll G$ is needed to achieve high
%fidelity. Otherwise, the amplitude of the state decays significantly before
%the state is transferred to another cavity. This is substantially different
%from the itinerant state transfer, which is possible even in the weak
%coupling regime as long as the cooperativity is high.

%If the coupling is strong enough,

In order to quantify the performance of a transfer protocol, we consider the
Ulhmann fidelity $F$~\cite{Uhlmann1976}.
\begin{equation}
F\equiv \left( \mathrm{Tr}[(\sqrt{\hrho _\text{i}}\hrho _\text{f}\sqrt{\hrho _\text{i}}%
)^{1/2}]\right) ^{2}.
    \label{eq:FidelityDefn}
\end{equation}%
where $\hrho _\text{i}$ ($\hrho _\text{f}$) denote the density matrix of initial state to be transferred (the final state in cavity 2). Note that we will optimize the fidelity over simple rotations in phase space (so that if $\hrho _\text{f}$ is a rotated
version of $\hrho _\text{i}$, $F=1$).

We will begin by considering the simple case of transferring a pure Gaussian state between the two
cavities (i.e.~a pure state whose Wigner function is Gaussian).  For such states, simple analytic expressions can be derived
for the transfer fidelity~\cite{Wang2012,Wallquist2010}, allowing quantitative insight into the various physical processes which degrade transfer.
Starting with an initial pure Gaussian state in cavity 1,
% and using the fact that due to linearity, the system remains in a general Gaussian state at all time,
 the transfer fidelity $F$ takes the general form (for details, see \ref{app:gaussianfidelity}):
\begin{equation}
F=\frac{1}{1+\bar{n}_{\mathrm{h}}}\exp \left( -\frac{\lambda ^{2}}{1+\bar{n}%
_{\mathrm{h}}}\right) .  \label{fg}
\end{equation}%
$F$ depends on just two positive-definite parameters $\bar{n}_{%
\mathrm{h}}$ and $\lambda $.  $\bar{n}_{\mathrm{h}}$ can be regarded as an
effective number of thermal quanta which quantifies the heating of the state
during the transfer protocol.  This heating can arise both from noise emanating from the cavity and mechanical
dissipative baths, as well as from any initial thermal population in the mechanical resonator.
In contrast $\lambda$ characterizes the deleterious effects of amplitude decay during the state transfer protocol, i.e. the decay of the average
of $\langle \hat{d} \rangle$.  Efficient
transfer requires minimizing both effects.

\begin{figure}[tbp]
\begin{center}
\includegraphics[width= 1 \columnwidth]{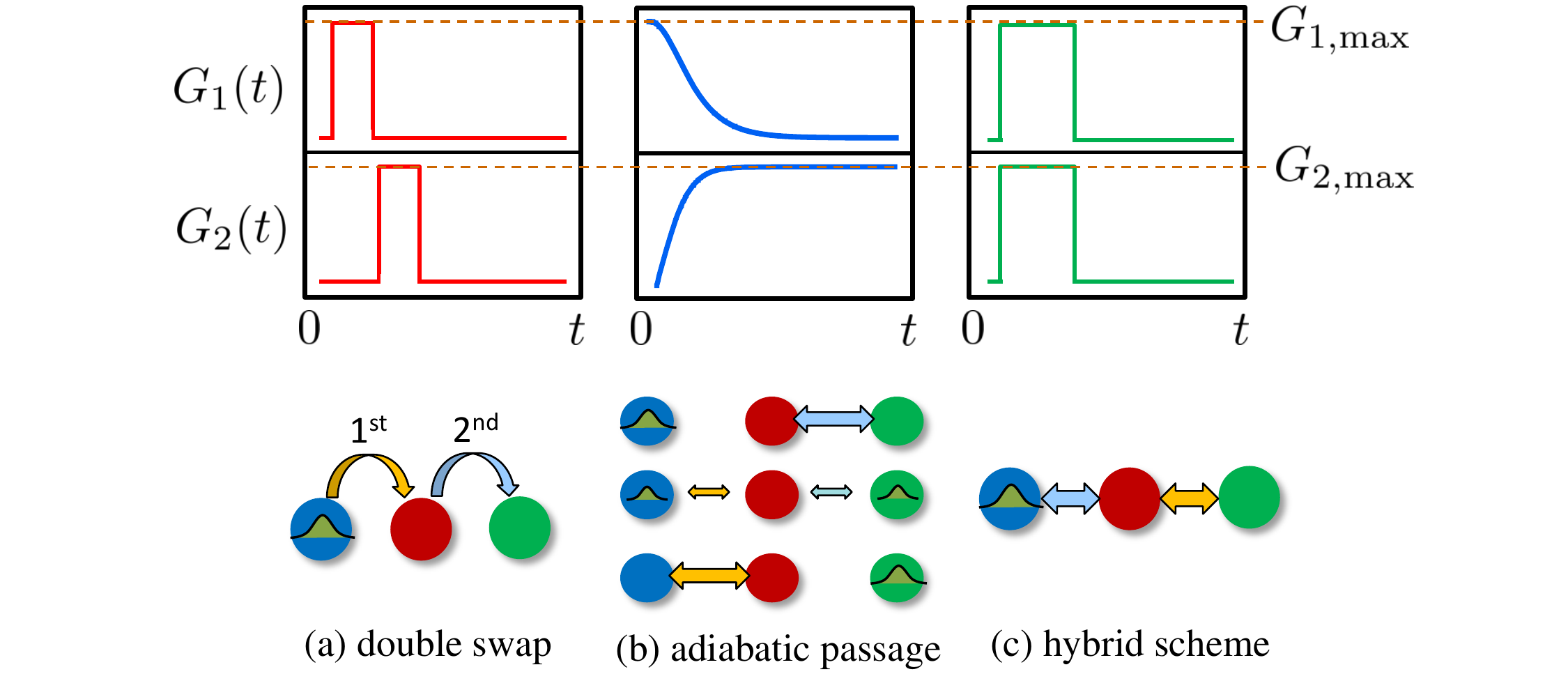}
\end{center}
\caption{(Color online) Schematics for the three intra-cavity state transfer schemes considered in this paper.  Top panel:
the curves in the boxes show the required time-dependences of the two optomechanical couplings in each scheme:
(a) double-swap (b) adiabatic passage, and (c) hybrid scheme. $G_i$ ($i=1,2$) is the optomechanical coupling to cavity $i$.  The brown dashed line denotes the maximum value of the coupling strength $G_{i,\max}$. Lower panel: schematics illustrating how the transfer works in each scheme.  Here the blue, red and green dots represent (respectively) cavity 1, the mechanical resonator, and cavity 2;  the wave packet in cavity 1 represents the state to be transferred.  Arrows indicate the optomechanical couplings, with the magnitude indicating their magnitude.}
\label{fig:pulse}
\end{figure}

The values of $\bar{n}_{\rm h}$ and $\lambda$ depend both on the nature of the initial Gaussian state, and
on the precise nature of the time evolution associated with the transfer protocol.  The 3 possible protocols correspond to different
time-dependences of the optomechanical couplings $G_1(t)$ and $G_2(t)$, as shown in Fig.~\ref{fig:pulse}. In the following subsections, we will
discuss the behavior of $\bar{n}_{\rm h}, \lambda$ and the fidelity $F$ for each of the three intra-cavity transfer schemes.

\subsection{Double-swap protocol}

\label{subsec:doubleswap}

As already discussed in the introduction, the most straightforward manner to exploit
Eq.~(\ref{ham}) for state transfer is via two sequential swap operations (see Fig.~\ref{fig:pulse}(a)).
%  Cavity 1 and the mechanical resonator
%are first made to interact so that they exchange their states; then cavity 2 and the mechanical resonator are made to interact and exchange their
%states (see Fig.~\ref{fig:pulse}a).
As this scheme only involves at most two modes interacting simultaneously, the mechanically-dark mode introduce earlier plays no role
in its physics.  In order to provide a basis for comparison for the dark-mode schemes that follow, we derive both the fidelity for this scheme, as well as discuss why this scheme requires the mechanical resonator to be initially prepared in its ground state.
%This general idea has been widely exploited in cavity QED~\cite{Zeng1994} as
%well as in optomechanical system \cite{Parkins1999,Tian2010,Wang2012}. Here
%we give a detailed description on it and the analytical result of the
%transfer fidelity based on this approach.

Using the Hamiltonian in Eq.~(\ref{ham}), the double-swap protocol requires that we first only
turn on the interaction between cavity 1 and the mechanical resonator (i.e. $G_{1}\left( t\right) =G_{1,\max }$ and $%
G_{2}\left( t\right) =0$ as shown in Fig.~\ref{fig:pulse}(a)).  In the absence of dissipation, the time
evolution of cavity 1 and mechanical mode reads
\begin{eqnarray}
\hat{a}\left( t\right) &=&e^{-i\omega _{M}t}\left( \hat{a}\left( 0\right)
\cos (G_{1,\max}t)-i\hat{d}_{1}\left( 0\right) \sin (G_{1,\max}t)\right)  \notag \\
\hat{d}_{1}\left( t\right) &=&e^{-i\omega _{M}t}\left( \hat{d}_{1}\left(
0\right) \cos (G_{1,\max}t)-i\hat{a}\left( 0\right) \sin (G_{1,\max}t)\right)
    \label{eq:swap}
\end{eqnarray}%
It thus follows that at a time $t_{1\mathrm{s%
}}=\pi /(2G_{1,\max })$, cavity 1 and the mechanical resonator will have perfectly exchanged their states (up to a trivial phase
factors) \cite{Parkins1999}:
\begin{equation}
    \hat{a} \left( t_{1\mathrm{s}}\right) =e^{-i\theta (t_{1\mathrm{s}})}
    \hat{d}_{1}\left( 0\right), \, \,\,
     \hat{d}_{1}\left( t_{1\mathrm{s}}\right) =e^{-i\theta
(t_{1\mathrm{s}})}\hat{a}\left( 0\right)
    \label{eq:PerfectSwap}
\end{equation}%
with $\theta (t)=\omega _{M}t+\pi /2$.
As the initial cavity-1 state is now in the mechanical resonator, one next turns off $G_1$ and turns on $G_2= G_{2,\max }$
for a time $t_{2 \rm{s}} = \pi / 2 G_{2,\max }$  to
similarly exchange the mechanical and cavity-2 states (see Fig.~\ref{fig:pulse}(a)).  At the end of the operation,
\begin{equation}
\hat{d}_{1}(t_{1\mathrm{s}}+t_{2\mathrm{s}})=e^{-i\theta (t_{1\mathrm{s}%
}+t_{2\mathrm{s}})}\hat{d}_{2}(0),\text{ \ }\hat{d}_{2}(t_{1\mathrm{s}}+t_{2%
\mathrm{s}})=e^{-i\theta (t_{1\mathrm{s}}+t_{2\mathrm{s}})}\hat{d}_{1}(0)
\end{equation}
Thus, without dissipation, one can perfectly transfer an arbitrary cavity-1 state to cavity-2.

\begin{figure}[tbp]
\begin{center}
\includegraphics[width= 1 \columnwidth]{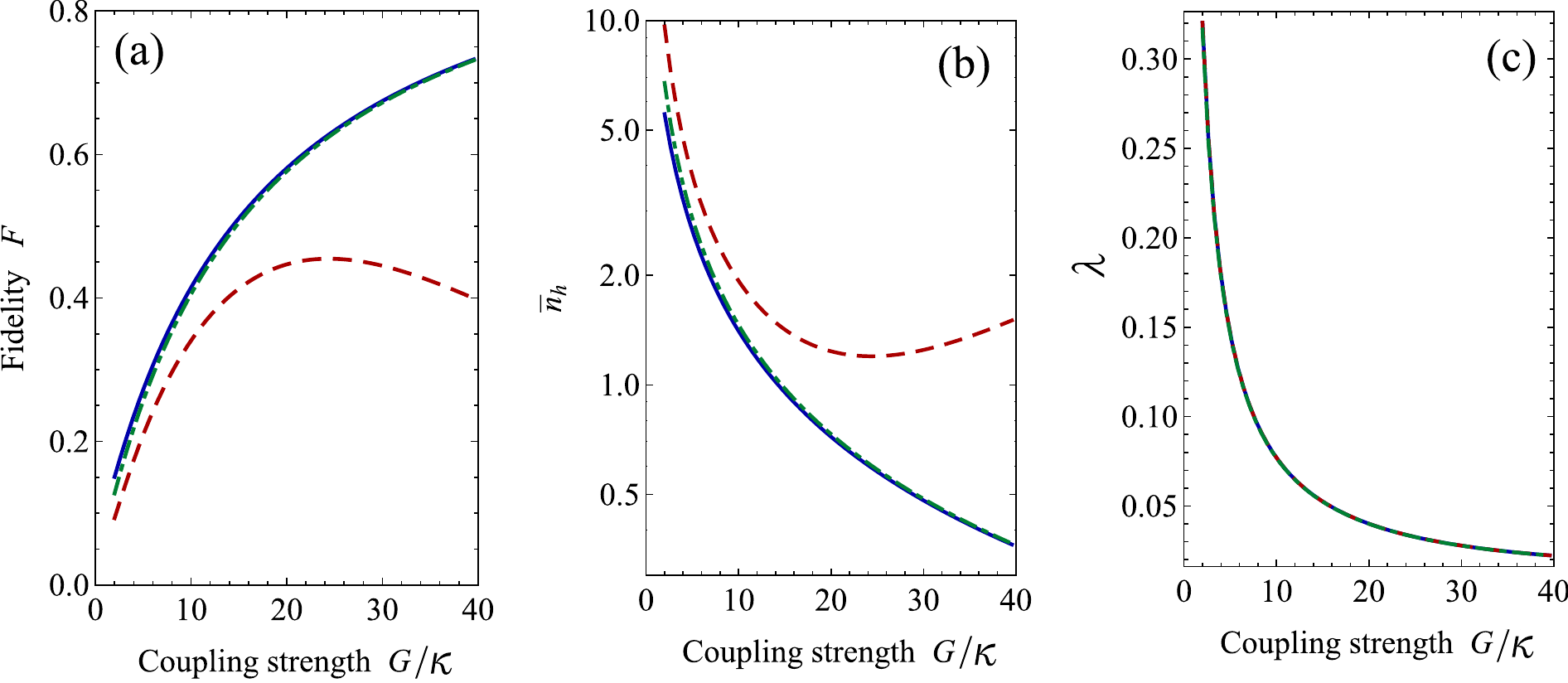}
\end{center}
\caption{(Color online) The performance of the double-swap scheme versus
coupling strength $G = G_{1,\max} = G_{2,\max}$, for the transfer of a coherent state having amplitude $\alpha = 1$.
Shown are (a) transfer
fidelity, (b) heating parameter $\bar{n}_\text{h}$ (note that a logarithmic scale is used), (c) damping parameter $\lambda$.
$\bar{n}_{\rm h}$ and $\lambda$ are defined after Eq.~(\ref{fg}).
We have fixed the bath temperature to be $T=1.5$~K,  (which would be compatible with the use of a superconducting microwave cavity fabricated from niobium),
and taken $ \kappa = (\kappa_1 + \kappa_2) / 2 = 2  \pi \times 50$~KHz (which is appropriate to the system envisaged in Ref.~\cite{Regal2011}).
The blue solid lines correspond to $ \kappa_1= \kappa_2$ and a mechanical resonator which is initially prepared
in its ground state; the red dashed lines correspond to
$ \kappa_1= \kappa_2$ but without any mechanical pre-cooling
(i.e.~the mechanical resonator is initially in thermal equilibrium at the bath temperature).
The green dashed-dotted line corresponds to the case where the mechanical resonator is precooled, but $\kappa_2 = 4 \kappa_1$.
Note that all lines coincide in plot (c), as the amplitude decay parameter $\lambda$ is insensitive to asymmetry
in the cavity damping, and to mechanical pre-cooling.
The other parameters are  $ \gamma=2  \pi \times 1$~KHz, $ \omega_M =-\Delta_i=2 \pi \times 10$~MHz. Cavity 1 (2)
is a microwave (optical) cavity: $\Omega_1/ 2  \pi =10$~GHz
($\Omega_2 / 2  \pi = 100$~THz). The transfer time is taken to be the optimal time
$t_\text{s}$ as defined in the text.}
\label{fig:doubleswap}
\end{figure}

In the presence of dissipation, one might expect that high fidelity transfer is still possible if the total transfer time
$t_{1 \rm{s}} + t_{2 \rm{s}}$ is much shorter than the relevant dissipative time scales.  This expectation can be easily quantified
in the case of Gaussian state transfer, using the general expression for the transfer fidelity $F$ given in Eq.~(\ref{fg}).  Not surprisingly,
one finds that each cavity must achieve the many-photon strong coupling condition $\kappa _{i}\ll G_{i,\max }$; otherwise, the decay of the state
in the two cavities during the transfer makes high fidelity impossible.  In this strong coupling regime, the heating parameter $\bar{n}_h$ and amplitude-decay parameter $\lambda$ that determine the fidelity take particularly simple forms.  The full expression for an arbitrary squeezed state (along with calculation details)
is presented in~\ref{app:doubleswap}.  Here, we present the results for the strong coupling case with a coherent state $|\alpha\rangle$ as the initial state. We also assume the
mechanical resonator has initially been prepared in the ground state. One finds that to leading order in $(\kappa+\gamma)/G$:
\begin{eqnarray}
\bar{n}_{\mathrm{h}} &=&\sum_{i}\frac{\gamma N_{M}+\kappa _{i}N_{i}}{2}t_{i\text{s}}\approx \gamma N_{M}\frac{\pi }{2G},  \notag \\
\lambda &=&\left\vert \alpha \right\vert \sum_{i}\frac{\kappa _{i}+\gamma }{4%
}t_{i\text{s}}\approx \left\vert \alpha \right\vert \frac{\kappa }{2}\frac{\pi }{2G}.  \label{ds}
\end{eqnarray}%
The effective number of thermal quanta $\bar{n}_{\rm h}$ generated during each time interval $t_{i \rm{s}}$ is simply given by the
duration of the interval times the average heating rate $\left( \gamma N_{M}+\kappa_i N_{i}\right) /2$.  Similarly,
the amplitude decay $\lambda$ in each interval is just the duration of the interval, times the amplitude of the initial state, times the average amplitude decay rate
$(\kappa _{i}+\gamma )/4$.  The last approximation in each equation above corresponds to the limit of
symmetric cavity parameters $\kappa _{i}=\kappa\gg \gamma $, $G_{i}=G$ and $N_{i}\approx 0$.
% Not surprisingly, if one does
%not have strong coupling $G_{i,\max }\ll \kappa _{i}$, the parameter $%
%\lambda \gg 1$, indicating that high-fidelity transfer is not possible as
%the state decays away before it can be swapped.

The above result implies that for mechanical heating to not be a problem, one needs that the optomechanical couplings $G$ not only be larger than $\kappa$, but also larger than the mechanical thermal decoherence rate $\gamma N_M$. This regime was recently realized in experiment \cite{Verhagen2012}. Plotted in Fig.~\ref{fig:doubleswap} are transfer fidelities corresponding to the transfer of a coherent state between a microwave cavity and an optical cavity, using numbers
appropriate to kind of setup envisaged in Ref.~\cite{Regal2011}; the behavior of the parameters $\bar{n}_{\rm h}$ and $\lambda$ are also shown over a range of
parameters. We have taken a relatively high bath temperature of $1.5$ Kelvin, as from an experimental point of view, it would be highly advantageous to not have to cool the optical cavity to tens of mK.  We note that such a high operating temperature is compatible with the use of a superconducting microwave resonator fabricated from niobium.
For such a high temperature, one clearly sees that the double-swap scheme is only able to yield good fidelities at extremely strong values of the optomechanical coupling $G$.

Degradation due to mechanical heating is even more severe when considering the transfer of
squeezed states (c.f. Eq.~(\ref{eq:FullSqueezednh})): this enhanced sensitivity is due to the state having narrower features in phase space.  As shown in Eq.~(\ref{nsqu}), when transferring a squeezed vacuum state, the general expression of Eq.~(\ref{fg}) for the fidelity still holds,
but the heating factor $\bar{n}_h$ is enhanced {\it exponentially} compared to the coherent-state case.  This is a direct consequence of the initial state having extremely narrow features in phase space; it suggests that the double-swap scheme will be especially poor in transferring such states (in comparison to schemes that attempt to suppress mechanical heating).
%; this is demonstrated in Fig. YY.
A similar conclusion can be expected when considering the transfer of non-classical states (e.g.~a Fock state), which also have fine structures in phase space.

We thus see that while the double-swap protocol works well at extremely low temperatures, it is already problematic at moderate values of mechanical bath temperature.
Given the experimental advantages of being able to use such temperatures, there is ample motivation to consider alternate transfer protocols which are more robust against the effects of mechanical thermal noise; we do this in the following subsections.

\begin{figure}[tbp]
\begin{center}
\includegraphics[width= 0.95 \columnwidth]{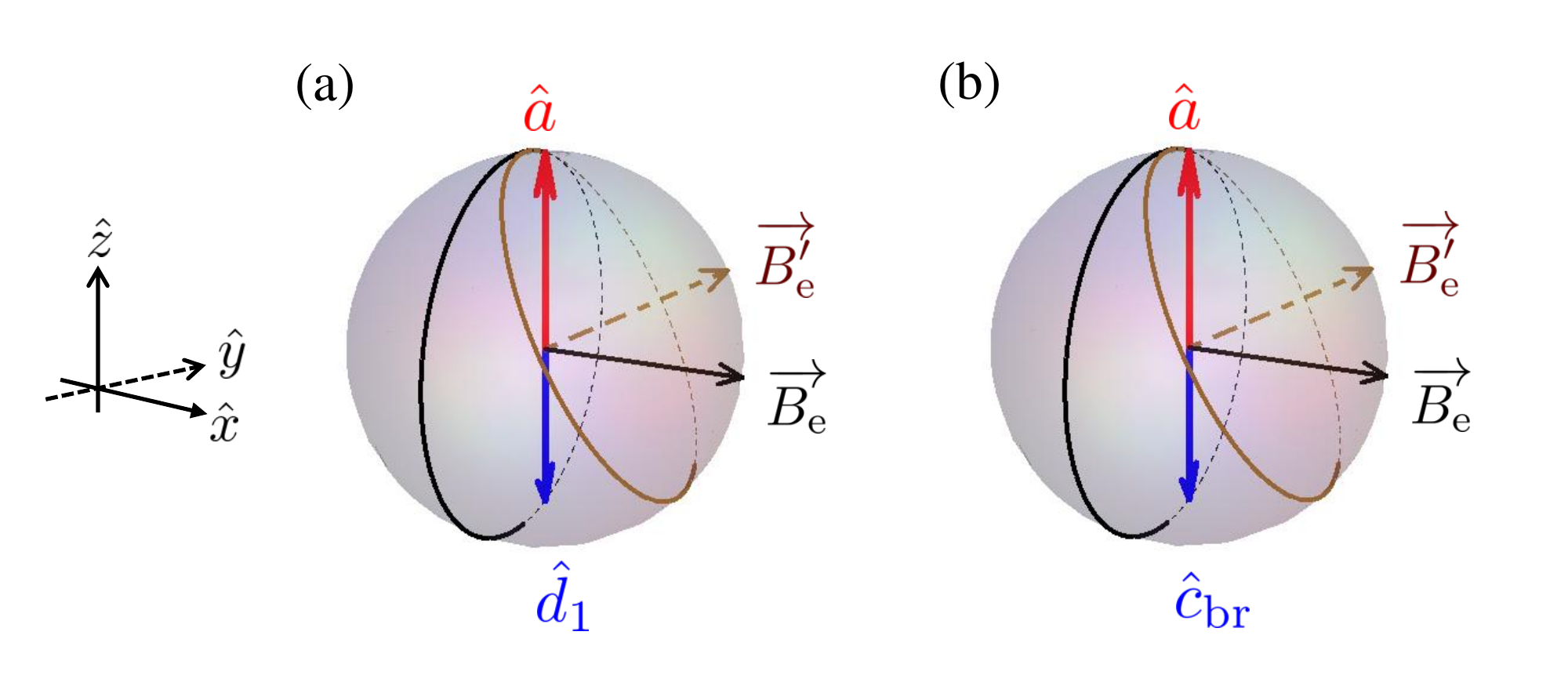}
\end{center}
\caption{(Color online) (a) Bloch sphere representation of a two-cavity swap operation.
Each swap operation in the double-swap transfer scheme (c.f.~Sec.\ref{subsec:doubleswap})
can be understood in analogy to the evolution of a two-level
system in a magnetic field; one takes the mechanical mode operator $\hat{a}$ and cavity mode operator $\hat{d}_1$ to correspond to
the $\sigma_z$ eigenstates of the TLS.  Without dissipation, each swap operation then corresponds to a $\pi$ rotation (as indicated by the blue circle on the sphere) around the
effective magnetic field $\overrightarrow{B_\text{e}}$, which is oriented along the $x$ axis and is proportional to
the optomechanical coupling $G_{i,\max}$.  As discussed in the text, including dissipation changes this picture.  The large asymmetry between the cavity damping $\kappa$ and mechanical damping $\gamma$ tilts the effective magnetic
field towards $z$-direction to $\overrightarrow{B^{\prime}_\text{e}}$.  A rotation about this tilted field now cannot
perfectly rotate $\hat{a}$ into $\hat{d}_1$ (as shown by the brown circle on the sphere):  a perfect swap is thus not possible.  This explains why the the double-swap scheme is so susceptible to an initial mechanical thermal population.
(b) A similar Bloch sphere representation can be used to understand the hybrid transfer scheme of Sec.~\ref{subsec:hybrid};
now however the $\sigma_z$ eigenstates of our effective TLS correspond to the mechanical mode and the ``bright"
cavity mode $\hat{c}_{\rm br}$ (c.f.~Eq.~(\ref{brm})), and the effective rotation that is needed is a $2 \pi$ rotation (see discussion
after Eq.~(\ref{eq:HybridTLS})).  As such a rotation can be performed even with a magnetic field that is not in the equatorial plane,
the hybrid scheme is insensitive to the initial mechanical population.}
\label{fig:bloch}
\end{figure}

Before proceeding, we consider the importance of precooling to the double-swap scheme.
 In Eq.~(\ref{ds}), we have assumed the optimal situation where the
mechanical resonator is initially in its ground state~\cite{Tian2010,Regal2011,Wang2012}, which could be achieved by initially swapping a cavity ground state into the mechanics,
or by using conventional cavity cooling \cite{Marquardt2007}.
If in contrast one does not initially cool the mechanics to the ground state, its initial thermal
population $N_{M,0}$ will make a significant additional contribution to
the heating parameter $\bar{n}_{\mathrm{h}}$.  One finds:
\begin{equation}
    \bar{n}_{\mathrm{h}}=\sum_{i}\frac{\gamma N_{M}+\kappa _{i}N_{i}}{2}t_{i\text{s}} +
    \left( \frac{\kappa _{1}-\gamma }{G_{1}}\right)
N_{M,0}  \text{.}  \label{nwoc1}
\end{equation}%
One might naively think that the initial thermal population of the mechanical resonator should be irrelevant, as it will just be swapped to cavity-1 in the initial swap operation of the protocol.  However, this is not the case due to cavity dissipation.  We can obtain a simple understanding of this by making an analogy between
our system and the evolution of a quantum two-level system (TLS) in a magnetic field.  Including only the damping effects of dissipation, and letting  $\vec{v} = \left(\hat{a}(t), \hat{d}_1(t) \right)$, the Heisenberg equation of motion for the cavity-1 and mechanical mode operators during the first swap pulse takes the form
\begin{equation}
    \frac{d}{dt} \vec{v} =
        -i \omega_M  \vec{v} - \frac{\kappa_1+\gamma}{2}\vec{v}
        -i G_{1,\max} \left(
            \begin{array}{cc}
                0   &   1   \\
                1   &   0
            \end{array}
        \right) \vec{v}
        - \frac{(\kappa_1-\gamma)}{2} \left(
            \begin{array}{cc}
                1   &   0   \\
                0   &   -1
            \end{array}
        \right) \vec{v}
        \label{eq:TLSanalogy}
\end{equation}

If we now interpret $\vec{v}$ as a spinor describing a quantum two-level system written in a basis of $\sigma_z$ eigenstates, then for $\kappa_1 = \gamma = 0$, the above equation corresponds to the precession of the TLS in a magnetic field of strength $\propto G_{1,\max}$ oriented in the $x$ direction.  The perfect swap
of cavity and mechanical states occurring at a time $t = t_{1\rm{s}}$ (c.f. Eq.~(\ref{eq:PerfectSwap}))
thus corresponds directly to a $\pi$ rotation of the effective TLS in the  Bloch sphere, as sketched in
Fig.~\ref{fig:bloch}(a) with the blue circle. Turning now to the dissipative terms in Eq.~(\ref{eq:TLSanalogy}), we see that there are two effects.  The first is an overall exponential
decay induced by the average damping rate $(\kappa_1+ \gamma)/2$; this corresponds to the second term on the RHS of Eq.~(\ref{eq:TLSanalogy}).  This decay on its
own would still allow for a perfect swap, in that $\hat{a}(t_{1 \rm{s}})$ would only depend on $\hat{d}_1(0)$ and not $\hat{a}(0)$.  In contrast, the last term on the RHS of Eq.~(\ref{eq:TLSanalogy}) (proportional to $(\kappa_1 - \gamma)/2$) is like an effective $z$-magnetic field acting on the TLS (albeit with an imaginary magnitude).  This terms makes a perfect swap impossible, as the effective magnetic field acting on our TLS is no longer purely transverse, i.e., in Fig.~\ref{fig:bloch}(a), the rotation axis is changed from $\overrightarrow{B_\text{e}}$ to $\overrightarrow{B'_\text{e}}$. It is this term that makes the double-swap protocol extremely susceptible to any initial thermal population in the mechanics:  this population is not fully swapped away in the first swap operation, and thus ends up being transferred to the final state of cavity 2.  As shown in Fig.~\ref{fig:doubleswap} by the dashed line, if one does not precool, the transfer fidelity of the double-swap scheme is greatly suppressed.  As we will see in Sec.~\ref{subsec:hybrid}, there does exist a transfer scheme (the hybrid scheme) which {\it does not} require any initial precooling of the mechanical resonator.

\subsection{Adiabatic passage}

\label{subsec:adiabatic}

%%%%%%%%%%%%%%%%%%%%%%%%%%%%
\begin{figure}[tbp]
\begin{center}
\includegraphics[width=1 \columnwidth]{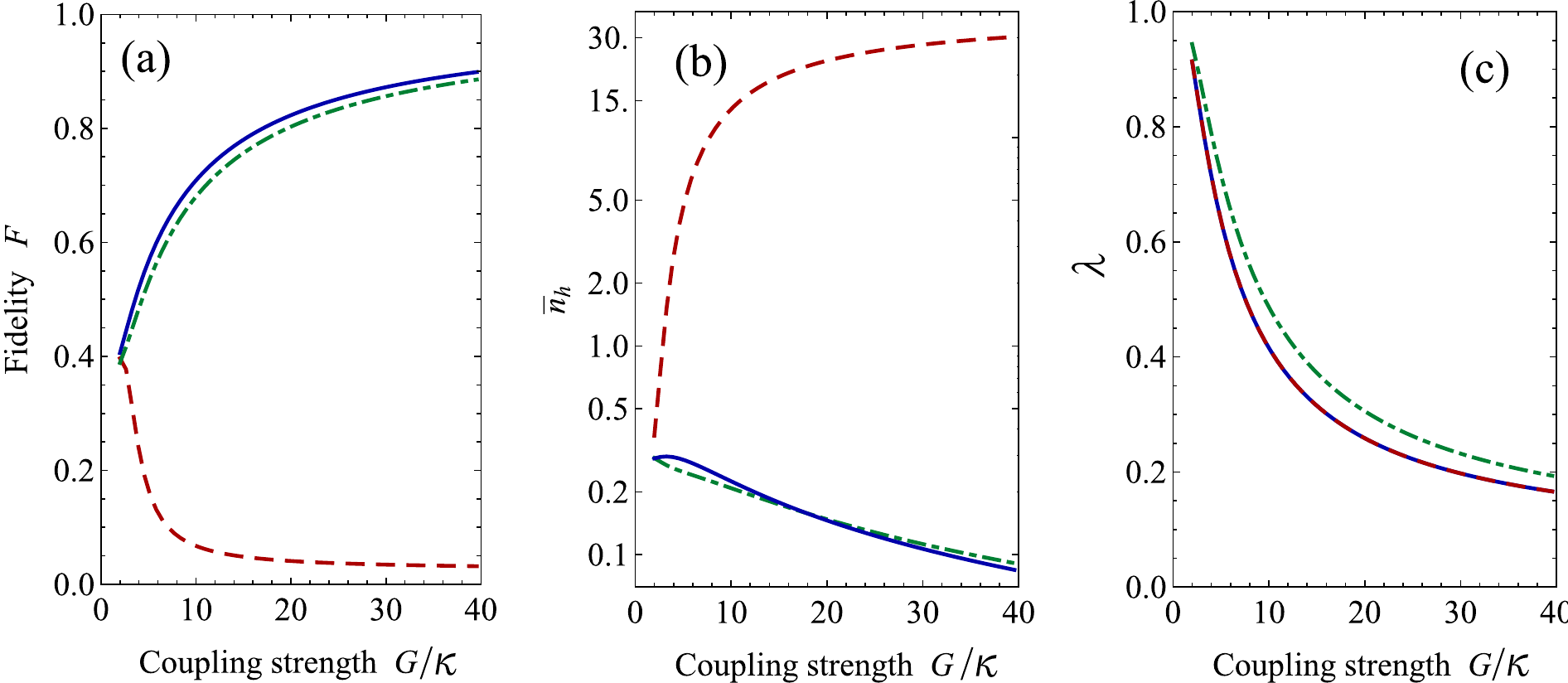}
\end{center}
\caption{(Color online) The performance of the adiabatic passage scheme
versus the maximum coupling strength $G=G_{1,\max}=G_{2,\max}$ for the transfer of a coherent state $| \alpha=1\rangle$,
using parameters is identical to Fig.~\ref{fig:doubleswap}.  Shown are:
(a) Fidelity, (b) the heating parameter $\bar{n}_\text{h}$ (notice that a logarithmic scale is used in this plot), (c) the damping
parameter $ \lambda$.
The blue solid lines correspond to $\kappa_1 = \kappa_2$ and a mechanical resonator that is initially
in the ground state; the red dashed line corresponds to the same without precooling.
The greed dashed-dotted lines correspond to a pre-cooled mechanical resonator but unequal cavity damping
$ \kappa_2=4 \kappa_1$ and $\kappa_1+\kappa_2=2\kappa$.
Even though asymmetrical cavity damping destroys the perfect protection of the mechanically-dark mode from
mechanical dissipation, it only results in a small decrease of the fidelity of the adiabatic transfer protocol; see text for details.
%As can be seen, the fidelity for the case with unequal
%cavity damping rate is slightly lower. But this is mainly due to the cavity damping instead of extra heating ($\bar{n}_\text{h}$ is barely
%changed. See the text for explanations).
%The cavity damping is stronger for $\kappa_1\neq\kappa_2$ because here we choose $\kappa_2$ to be larger; and among the total operation time, $G_2$ is switched on much longer (see Fig.~\ref{fig:pulse}(b)).
%In fact, here if we choose $\kappa_1=4\kappa_1$, the fidelity of unequal $\kappa$ is even higher than the equal $\kappa$ case.
For each value of the coupling,
we have optimized the speed $\beta$ and final time $t_f$ of the protocol to obtain a maximum fidelity (see main text).}
%The other parameters are $T=1.5$~K,
%$ \gamma=2  \pi \times 1$~KHz, $ \omega_M =-\Delta_i=2%
% \pi\times 10$~MHz. Cavity 1 (2) is a microwave (optical) cavity: $%
%\Omega_1/ 2  \pi =10$~GHz ($\Omega_2 / 2  \pi = 100$~THz), $%
% \kappa_1 = 2  \pi \times 50$~KHz. The transfer time is taken
%to be the optimal time $t_\text{opt}$ as defined in the text.}
\label{fig:adiabatic}
\end{figure}
%%%%%%%%%%%%%%%%%%%%%%%%%%%%

The double-swap scheme of the previous section only involves two modes interacting with one another at any given time,
and hence could not take advantage of the full structure of our three-mode (two cavities, one mechanical resonator) system.  Here, we
consider an alternative scheme (first discussed in Refs. \cite{Wang2012,Tian2012})
which explicitly makes use of the ``mechanically-dark" mode $\hat{c}_{\mathrm{dk}}$ introduced in Sec.~\ref{subsec:darkmode}
(c.f.~Eq. (\ref{dkm})), a mode which is not coupled directly to the mechanical resonator and hence protected against mechanical dissipation.  The basic idea
is to use this dark mode to perform the state transfer from cavity 1 to 2 {\it entirely} via the dark mode.  This is done by adiabatically modulating the optomechanical couplings $G_1$, $G_2$ in time,
so that at $t=0$ the dark mode is equal to the cavity-1 mode, and at the end of the protocol $t=t_f$ it
is equal to the cavity-2 mode.  This involves using a pulse sequence where $G_1$ and $G_2$ are modulated in time as shown in Fig.~\ref{fig:pulse}(b).
$G_{1}\left( t\right) $ is increased from $0$ to $G_{1,\max}$%
, while $G_{2}\left( t\right) $ is simultaneously decreased from $G_{2,\max}$ to $0$.  For the transfer to be perfectly adiabatic, one requires that the transfer speed be slow compared to the energetic gap separating the dark mode and the two coupled modes
$\hat{c}_{\pm}$ defined in Eq.~(\ref{cpm}).  This energetic gap is given by
$\sqrt{G_{1}^{2}+G_{2}^{2}}$ (c.f.~Eq.~(\ref{eq:omegapm})).  If one is indeed adiabatic, the dark mode will evolve during this protocol from being $-\hat{d}_{1}$ at $t=0$ to $\hat{d}_{2}$ at
the end of the protocol at a time $t=t_{f}$.

The adiabatic passage scheme described here is analogous to the well known
Stimulated Raman Adiabatic Passage (STIRAP) scheme employed in atomic physics (see Ref.~\cite{Bergmann1998} and reference therein).
In STIRAP schemes, one employs a dark-state of a 3-level atom, and also uses a similar ``counter-intuitive" pulse sequence to transfer an atomic population between two states.  In contrast, here we have an entire manifold of dark states (i.e.~any states in the subspace generated by $\hat{c}^\dagger_{\rm{dk}}$), and thus the potential to transfer an arbitrary cavity state using the adiabatic protocol.

To quantify the effectiveness and fidelity of this scheme, we again consider the transfer for a Gaussian state; the fidelity $F$ thus
takes the general form described by Eq.~(\ref{fg}).  While there are many possible ways to implement the adiabatic passage scheme, for simplicity, we use throughout the simple pulse shapes
\begin{equation}
G_{1}\left( t\right) =G\sin \left( \frac{\pi }{2}\tanh \beta t\right)
,G_{2}\left( t\right) =G\cos \left( \frac{\pi }{2}\tanh \beta t\right).
\end{equation}%
This form keeps the energy splitting between the dark and coupled modes constant (i.e. $G_{1}^{2}\left( t\right)
+G_{2}^{2}\left( t\right) =G^{2}$), a feature that has been argued to be optimal~\cite{Vasilev2009}.  The parameter $\beta$ represents the overall speed of the modulation; one would require $\beta \ll G$ to be in the purely adiabatic limit.

Shown in Fig.~\ref{fig:adiabatic} are results for fidelity, heating parameter $\bar{n}_h$ and amplitude decay parameter $\lambda$ for the adiabatic passage scheme, using parameters identical to Fig.~\ref{fig:doubleswap} (in particular, a bath temperature of 1.5 K).  For each data point, we numerically found the optimal values of the transfer speed $\beta$ and transfer time $t_f$, and use these to obtain the fidelity.  As expected, using the dark-mode leads to a dramatically
suppressed $\bar{n}_{\mathrm{h}}$ compared to the double swap scheme (i.e.~compare solid curves in Fig.~\ref{fig:doubleswap}(b)
and Fig.~\ref{fig:adiabatic}(b)).   If the mechanical bath temperature is sufficiently
high, the adiabatic transfer scheme yields far higher fidelities than the
double-swap scheme.  This behavior is shown explicitly in Fig.~\ref{fig:FidelityVersusT}, where the transfer
fidelity versus temperature for both these schemes are plotted.  We will quantify this advantage in more detail in
Sec. \ref{subsec:compare}, where we compare and contrast the performance of all
transfer schemes.

One might be puzzled by the results of Fig.~\ref{fig:adiabatic}(b), which shows that the effective heating parameter
$\bar{n}_{\rm h}$ is never exactly zero, even in the case $\kappa_1 = \kappa_2$,
where the dark mode is completely isolated from the mechanical resonator.  The reason is simple to understand:  because of the effects of cavity damping, the optimal
pulse speed $\beta$ in the adiabatic transfer protocol can never be zero, and hence for realistic cavity parameters,  one will never be in the perfectly adiabatic limit.  The result is some mixing into the non-adiabatic subspace, and hence some mechanically-induced heating.  If one were in the adiabatic limit $\beta \ll G$, mechanical noise would indeed not be a problem, but the slow transfer time would mean that the decay of the state in the two cavities would greatly degrade the fidelity.  Thus, the
two conflicting requirements of better
adiabaticity (to suppress mechanical heating) and faster operation time (to suppress cavity damping effects) result in an optimal, non-zero value of $%
\beta _{\text{opt}}$ and an optimal transfer time $t_{\text{opt}}$. High
fidelity transfer is only possible when $G\gg \beta \gg \kappa $, as it is only in this regime where one can both be reasonably adiabatic as well as reasonably fast compared to the cavity damping rate $\kappa$.  One also finds that the optimal value
$\beta _{\text{opt}}$ has a marked dependence on temperature, as the relative importance of being
adiabatic increases as the mechanical thermal noise increases.  The fact that one is not in the perfectly
adiabatic limit also leads to the requirement that the mechanical resonator to initially start in the ground state, as shown by the marked difference between the red-dashed and blue-solid lines in Fig.~\ref{fig:adiabatic} (a) and (b).
%The blue dots in
%Fig.~\ref{fig:phasediagram} are obtained with the modulation rate and the
%total transfer time optimized for each temperature.

The above discussion also explains why for realistic parameters, there is no stringent requirement that the cavity damping be symmetric,
i.e. $\kappa_1 = \kappa_2$; even highly asymmetric damping only slightly reduces the transfer fidelity  (see the green
dash-dot line in Fig.~\ref{fig:adiabatic}). As discussed in Sec.~\ref{subsec:darkmode}, asymmetric cavity damping mixes the cavity dark and bright modes, thus opening up the dark mode to mechanical noise. However, as the optimal transfer speed never corresponds to being in the perfectly adiabatic limit, the
additional exposure to the mechanical noise caused by unequal $\kappa $ has rather small contribution to the heating
parameter $\bar{n}_{\text{h}}$ above that stemming from non-adiabatic transitions.  The upshot is that the adiabatic transfer scheme can be useful even if one has rather dissimilar cavity dissipation rates.

%Finally, the fact that one is never in the perfectly adiabatic limit also explains why it is necessary to start the mechanical resonator in the ground state in the adiabatic transfer protocol.  If one had truly adiabatic evolution, the system would always be in the dark mode and would never see the mechanical mode or its initial thermal population.  However, as the transfer speed is finite, non-adiabatic transitions lead to sensitivity to the initial mechanical population; as a result, not precooling will
%significantly lower the
%transfer fidelity (see the red dashed lines in Fig.~\ref{fig:adiabatic}).
%
Finally, we note that in the perfectly adiabatic limit, the parameters $\bar{n}_\text{h}$ and $\lambda$ can be calculated analytically~\cite{Tian2012}.  However, for moderate $G / \kappa$ ratios, the optimal pulse shapes are not perfectly adiabatic, and hence these expressions are not particularly relevant.

%%%%%%%%%%%%%%%%%%%%%%%%%%%%%%%%%%%%%%%%%%%%%%%%%%%%%%%
%%%%%%%%%%%%%%%%%%%%%%%%%%%%%%%%%%%%%%%%%%%%%%%%%%%%%%%
%%%%%%%%%%%%%%%%%%%%%%%%%%%%%%%%%%%%%%%%%%%%%%%%%%%%%%%
\subsection{Hybrid transfer scheme}

\label{subsec:hybrid}

%%%%%%%%%%%%%%%%%%%%%%%%%%%%%%%%%%%%%%%%%%%%%%%%%%%%%%%
\begin{figure}[t]
\begin{center}
\includegraphics[width= 0.7 \columnwidth]{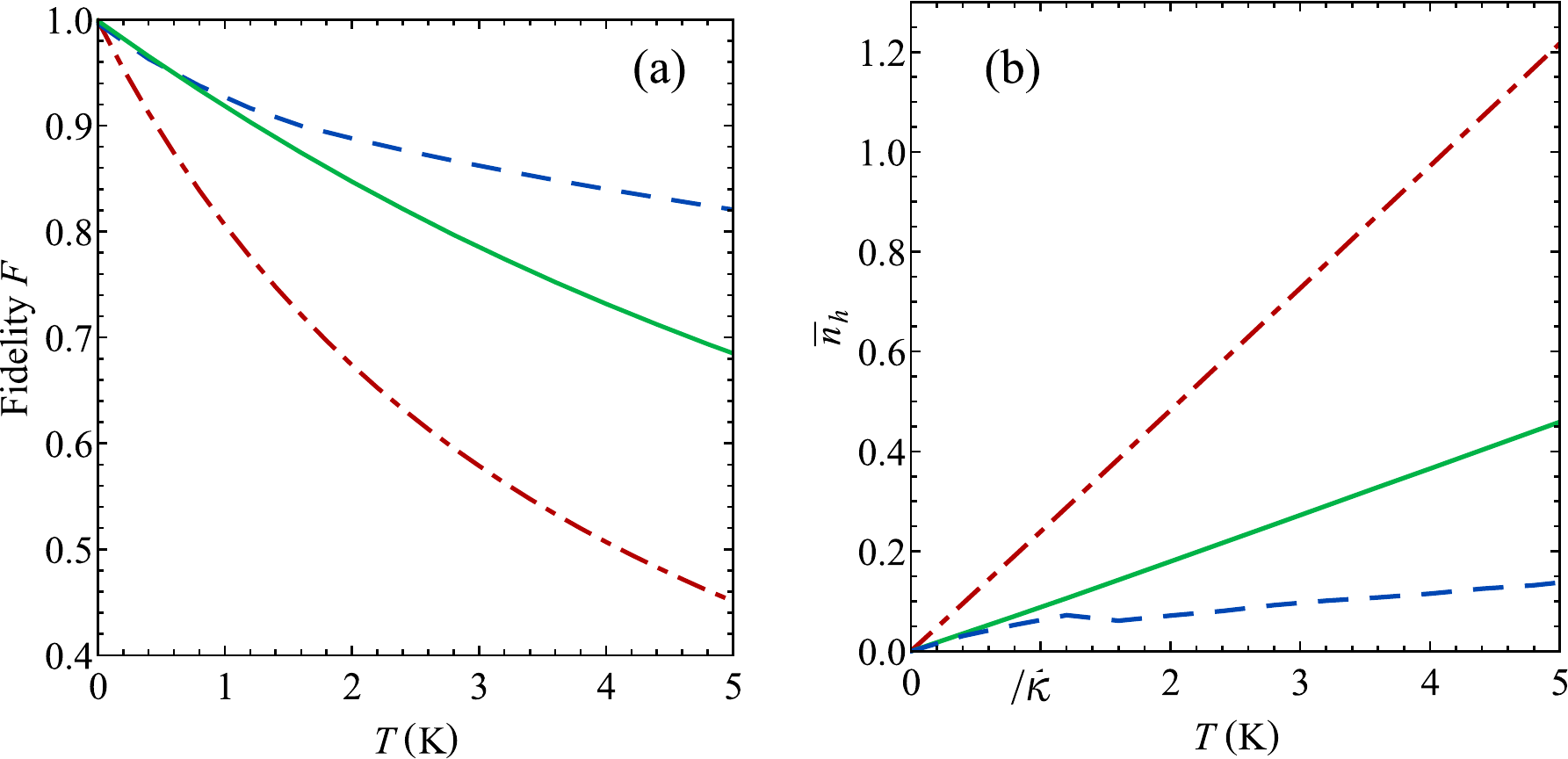}
\end{center}
\caption{(Color online) Comparison of the performance of the three intra-cavity transfer schemes for the transfer
of a coherent state of amplitude $\alpha = 1$, as a function of bath temperature.  We have taken
$G_{1,\max} = G_{2,\max} = 2 \pi \times 2$~MHz, the other parameters are identical to
Figs.~\ref{fig:doubleswap} (a).  Shown are (a) fidelity, and (b) heating parameter $n_{\rm h}$.  The blue dashed
curve is the adiabatic transfer protocol, the green solid curve the hybrid scheme, and the red dash-dotted curve the double-swap scheme.  As expected, the use of the mechanically-dark mode in the hybrid and adiabatic transfer schemes afford a much stronger
resilience against the effects of mechanical heating.}
\label{fig:FidelityVersusT}
\end{figure}
%%%%%%%%%%%%%%%%%%%%%%%%%%%%%%%%%%%%%%%%%%%%%%%%%%%%%%%

So far in this section, we have considered two very different approaches to state transfer,  each effective in
a very different limit.  The
double-swap scheme of Sec.~\ref{subsec:doubleswap}  does not involve the mechanically-dark mode in any way, and is hence more susceptible to mechanical thermal noise; however, it is relatively fast (the transfer time $t \propto 1/G$), and hence less susceptible to degradation due to cavity damping.  In contrast, the adiabatic passage scheme of
Sec.~\ref{subsec:adiabatic} attempts to {\it only} use the dark mode for state transfer.  The result is a strong resilience against mechanical noise.  However, the constraint of being near-adiabatic means that this scheme is slow (i.e.~the transfer time $t \gg 1/G$), and hence more susceptible (relative to the double-swap scheme) to the effects of cavity damping.

Given that both schemes have their relative merits, one might be
tempted to develop a strategy that incorporates the best features of both.
We present such a scheme here, the so-called ``hybrid" scheme, which utilizes {\it equally} the mechanically-dark
mode (c.f.~Eq.~(\ref{dkm}))  and mechanically-coupled cavity modes (c.f.~Eq.~(\ref{cpm})) for a state transfer between the two cavities.  As we will see, the fact that half the transfer is done
using the dark mode reduces the effects of mechanical thermal noise by a factor of two over the double-swap scheme.  However, unlike the adiabatic passage scheme, this protocol is relatively fast (i.e. the transfer time is $\sim 1/G$), and hence the effects of cavity damping are similar to what one would expect in the double-swap scheme.  As an added bonus, we also find that this hybrid scheme does not require any precooling of the mechanics:  high transfer fidelities are possible without having to initially cool the mechanical resonator to its ground state.  For typical experimental situations where both cavity damping and mechanical heating are of equal importance, this hybrid scheme  will in general yield the highest transfer fidelities (something we will make more clear in Sec.~\ref{subsec:compare}, where we compare the various intra-cavity state transfer protocols).
%
%
%
%
% That is a scheme based on the true $3$ mode system and employing
%both the dark mode $\hat{c}_{\text{\textrm{dk}}}$ (cf. Eq.~(\ref{cpm})) and
%mixed modes $\hat{c}_{\pm }$ (cf. Eq.~(\ref{dkm})), which is termed by us as
%\textquotedblleft hybrid scheme\textquotedblright . In this scheme, half of
%the transfer will be done via the dark mode (and hence will be immune to the
%mechanics), whereas half the transfer will involve actually putting the
%state in the mechanical resonator. The hybrid feature makes it more
%insensitive to mechanical heating than the double-swap scheme, while at the
%same time have a much faster operation time than the adiabatic scheme (and
%hence more resilience against cavity damping).

The hybrid scheme we have in mind is realized in an extremely simple fashion:  one simply turns on simultaneously both optomechanical couplings
$G_{1}$ and $G_{2}$ to equal magnitudes, as shown in Fig.~\ref{fig:pulse}(c).  It is simplest to consider the physics using the basis of dark and bright modes to describe the two cavities (c.f.~Eqs.~(\ref{brm}),(\ref{dkm})).  Taking $G_{1,\max} = G_{2,\max} = G$, one finds:
\begin{eqnarray}
    \hat{d}_1 & = & \frac{1}{\sqrt{2}} \left( \hat{c}_{\rm{br}} + \hat{c}_{\rm dk} \right)  \label{eq:c1Decomp}\\
    \hat{d}_2 & = & \frac{1}{\sqrt{2}} \left( \hat{c}_{\rm{br}} - \hat{c}_{\rm dk} \right)   \label{eq:c2Decomp}
\end{eqnarray}
The dark mode is uncoupled to the mechanical resonator, whereas the bright cavity mode and mechanical resonator evolve according to the Heisenberg equation of motion:
 \begin{equation}
    \frac{d}{dt} \vec{w} =
        \left(-i \omega_M - \frac{\kappa+\gamma}{2}\right) \vec{w}
        -i \sqrt{2} G \left(
            \begin{array}{cc}
                0   &   1   \\
                1   &   0
            \end{array}
        \right) \vec{w}
        - \frac{(\kappa-\gamma)}{2} \left(
            \begin{array}{cc}
                1   &   0   \\
                0   &   -1
            \end{array}
        \right) \vec{w}
        \label{eq:HybridTLS}
\end{equation}
% \begin{equation}
%   \frac{d}{dt} \vec{w} =
%       \left(-i \omega_M - \frac{\kappa+\gamma}{2}\right) \vec{w}
%       -i \sqrt{2} G \left(
%           \begin{array}{cc}
%               0   &   1   \\
%               1   &   0
%           \end{array}
%       \right) \vec{w}
%       - \frac{(\kappa-\gamma)}{2} \left(
%           \begin{array}{cc}
%               1   &   0   \\
%               0   &   -1
%           \end{array}
%       \right) \vec{w}
%       \label{eq:HybridTLS}
%\end{equation}
where $\vec{w} = (\hat{c}_{\rm br}, \hat{a})$, and we have taken $\kappa_i=\kappa$ and neglected noise terms for simplicity. Note the similarity between this equation and Eq.~(\ref{eq:TLSanalogy}) which describes the first step of the double-swap scheme.

Without dissipation (i.e. $\kappa _{i}=\gamma =0$), the situation here is analogous to the double-swap scheme:  the evolution of the two interacting modes is equivalent to that of a TLS in a transverse magnetic field.  In this case however, the $\sigma_z$-eigenstates of the effective TLS corresponds to the cavity bright mode and the mechanical mode.  From Eqs.~(\ref{eq:c1Decomp}),(\ref{eq:c2Decomp}), it follows that to swap the states of the two cavities, we simply need to change the sign of the bright mode relative to the dark mode.  This is easily achieved by performing a $2 \pi$ rotation of our effective TLS.  Thus, after a time $t= t_{\text{hs}}=\pi /(\sqrt{2}G)$, we find:
\begin{equation}
\hat{d}_{1}(t_{\text{hs}})=e^{-i\theta }\hat{d}_{2}\left( 0\right) ,\text{ \ }\hat{d}_{2}(t_{\text{hs}})=e^{-i\theta }\hat{d}_{1}\left( 0\right)
\end{equation}%
with $\theta =\omega _{M}t_{\text{hs}}+\pi $. Thus, similar to the
double-swap scheme, the state of the two cavities are swapped.

\begin{figure}[tbp]
\begin{center}
\includegraphics[width= 1 \columnwidth]{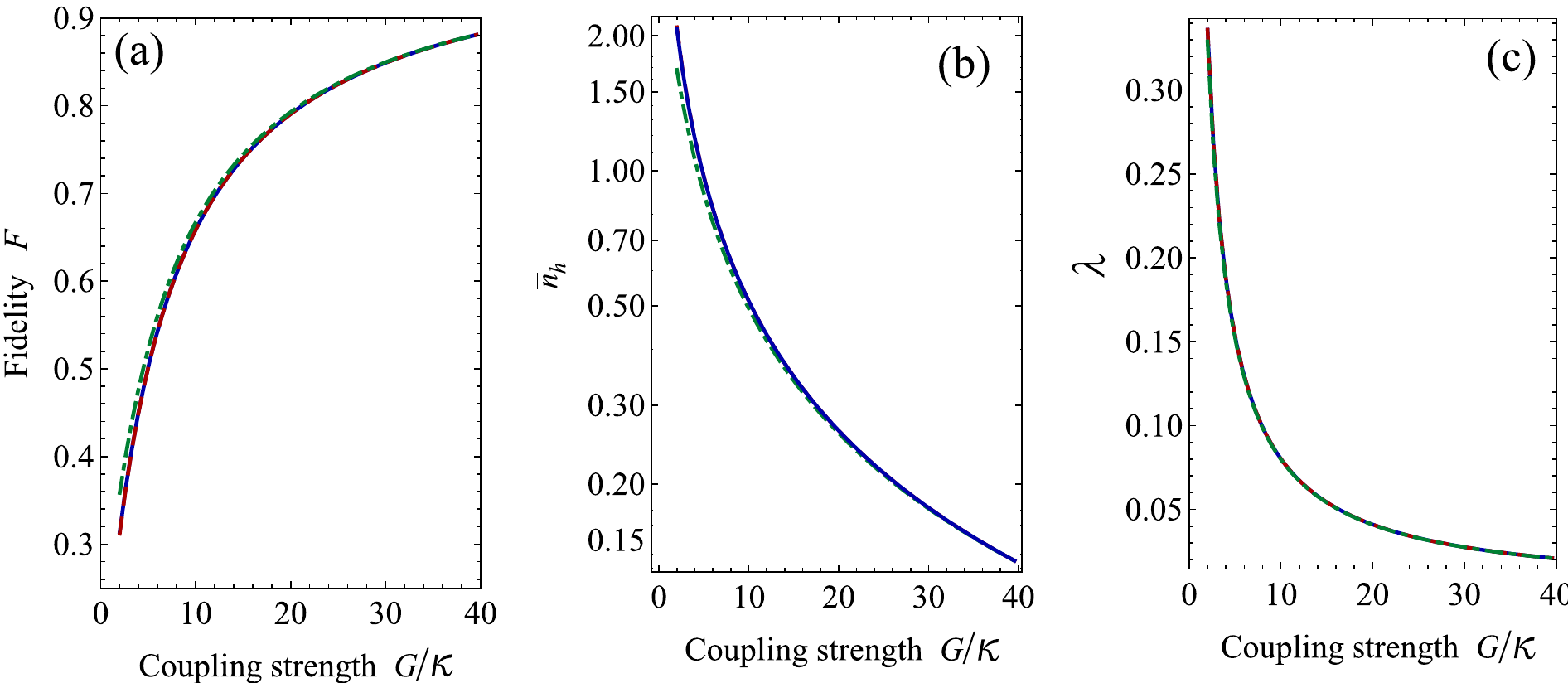}
\end{center}
\caption{(Color online) The performance of the hybrid scheme versus coupling
strength $G=G_{1,\max}=G_{2,\max}$ for the transfer of a coherent state of amplitude $\alpha = 1$, at a bath temperature $T = 1.5$~K;
parameters are the same as Figs.~\ref{fig:doubleswap} and \ref{fig:adiabatic}.
Shown are: (a) fidelity, (b) the heating parameter $\bar{n}_\text{h}$, (c) the damping parameter $\lambda$.
The blue solid line corresponds to $\kappa_1 = \kappa_2=\kappa$ and a mechanical resonator initially prepared in the ground state;
the red dashed line corresponds to the same, but without any mechanical precooling (the mechanics starts from a thermal equilibrium state).  As discussed in the main text, the hybrid scheme is almost completely insensitive to any initial thermal population in the mechanical resonator. The green dashed-dotted lines correspond
to the case with pre-cooling but unequal cavity damping $ \kappa_2=4\kappa_1$. The transfer time is taken to be the optimal
time $t_\text{hs}$ as defined in the text.}
\label{fig:hybrid}
\end{figure}

Unlike the double-swap process, this hybrid scheme involves both cavities being simultaneously coupled; this will have implications in terms of how the transfer is now affected by dissipation. On a heuristic level, the initial state in cavity $1$ corresponds to having an equal initial population of the cavity bright and dark modes.  Thus, half of the transfer is performed via the dark state, and we expect a better resilience against mechanical thermal noise than the double swap scheme. To quantify this, we again consider using the hybrid scheme to transfer a Gaussian state, including the full effects of cavity and mechanical dissipation.  The
fidelity $F$ thus takes the general form given in Eq.~(\ref{fg}) (see~\ref{app:hybrid} for details), with the heating and amplitude decay parameters in the strong coupling regime:
\begin{eqnarray}
\bar{n}_{\mathrm{h}} &=&\frac{\gamma N_{m}+3\kappa N_{c}}{4}t_{\mathrm{h}\text{s}}\approx \frac{\gamma N_{m}}{2\sqrt{2}}\frac{\pi }{2G},  \notag \\
\text{ }\lambda &=&\left\vert \alpha \right\vert \frac{\gamma +3\kappa }{8}
t_{\mathrm{h}\text{s}}\approx \left\vert \alpha \right\vert \frac{3\kappa }{4\sqrt{2}}\frac{\pi }{2G}  \label{hb}.
\end{eqnarray}%
For simplicity, we have taken the two cavities to be identical: $\kappa_1 = \kappa_2=\kappa$, $N_1 = N_2 = N_c$.
%
%where we have assumed the parameters of the two cavities equal $G_{1}=G_{2}=G $, $\kappa _{1}=\kappa _{2}=\kappa $, $\bar{\gamma}=\gamma
%\left( 2N_{m}+1\right) $, $\bar{\kappa}_{i}\approx \bar{\kappa}$, $N_{i}=N_{c}$.
The last approximation in each equation corresponds to the usual case where the mechanics dominates the heating, while the cavity dominates the amplitude decay.
Comparing the expression for $\bar{n}_{\rm h}$ with that of the the double swap case in Eq.~(\ref{ds}), we find that the effect of mechanical thermal decoherence is reduced here
by a factor of $2\sqrt{2}$.  This comes from the fact that half of the input state is transferred via the dark mode, and from the shorter transfer time in the hybrid scheme, i.e., $\pi / (\sqrt{2} G)$ versus $ \pi/ G$.  The effect of cavity damping is only slightly stronger than in the double swap case (by a factor $3\sqrt{2}/4\approx 1.06$).  Fig.~\ref{fig:FidelityVersusT} shows the transfer fidelity of a $\alpha=1$ coherent state versus temperature for both the hybrid and double-swap schemes, while Fig.~\ref{fig:hybrid} shows how the hybrid-scheme fidelity varies
with coupling; both demonstrate the greater resilience against thermal heating of the hybrid scheme over the double-swap scheme.

%%%%%%%%%%%%%%%%%%%%%%%%%%%%%%%%%%%%%%%%%%%%%%%%%%%%%%%
%%%%%%%%%%%%%%%%%%%%%%%%%%%%%%%%%%%%%%%%%%%%%%%%%%%%%%%

As already mentioned, an additional (and surprising) advantage of the hybrid scheme over both the
adiabatic and double-swap schemes is that it does not require the mechanical resonator to start in the ground state.  We can understand
this heuristically by again using our analogy to the precession of a TLS.  As we have explained, both the
double-swap and the hybrid schemes corresponds to effective rotations in the Bloch
sphere around a magnetic field $\mathbf{B}_{\text{e}}$ (see Fig.~\ref{fig:bloch}).  In the absence of dissipation, this
magnetic field is aligned along the $x$ axis and can be used to perform a perfect $\pi$ rotation (in the double-swap scheme) or
perfect $2 \pi$ rotation (in the hybrid scheme). Including mechanical and cavity dissipation induces new terms in the equations of motions, which
can be interpreted as an additional (unwanted) magnetic field along the $z$-direction (c.f.~Eqs.~(\ref{eq:TLSanalogy}) and~(\ref{eq:HybridTLS})).
As discussed, this misalignment of the magnetic field means that a perfect $\pi$ pulse is no longer possible, and hence in the double-swap scheme,
one is sensitive to the initial mechanical state. In contrast, in the hybrid scheme we need to perform a $2 \pi$ rotation.
This is possible {\it even if} the dissipation causes the effective magnetic field to be misaligned and not purely transverse. The result is that even with dissipation, the final cavity-2 state will remain completely independent of the initial mechanical thermal population. This is the reason that the performance with pre-cooling (blue solid line) and without pre-cooling (red dashed line) overlap as shown in Fig.~\ref{fig:hybrid}.

Finally, from Fig.~\ref{fig:hybrid}, we also see that the hybrid scheme is insensitive to the asymmetry of cavity damping (the green dashed line only deviates from the blue solid line slightly). This is because the thermal heating here is dominated by the the use of the coupled
cavity modes; the small heating of the dark mode that occurs when $\kappa_1 \neq \kappa_2$ is only a small correction to this, especially given the relatively short transfer time of the scheme (it is the fastest of the three schemes we consider).
%because the transfer time is short (shortest in all 3 schemes) and the transfer is already exposed partially to the mechanicaleating even if $\kappa_1=\kappa_2$.
%
%
%
% which is
%aligned with $x$ axis. Environmental dissipation leads to a deviation of the
%effective magnetic field towards $z$ direction, i.e. $\mathbf{B}_{\text{e}}$
%is changed into $\mathbf{B}_{\text{e}}^{\prime }$ in Fig.~\ref{fig:bloch}.
%Therefore, a perfect swap represented by a $\pi $ rotation from the south
%pole to the north pole cannot be accomplished. This leads to a dependence on
%the initial mechanical population as in Eq.~(\ref{nwoc1}). However, in the
%hybrid scheme, the rotation is $2\pi $ instead of $\pi $. A $2\pi $ rotation
%can always return to the original point regardless of the rotation axis.
%Therefore, as long as operation time is properly chosen to fulfill the $2\pi
%$ angle, the simple phase change needed in the hybrid scheme can be performed exactly, and the final
%cavity-2 state will not know about the initial mechanical state.  The heating
%of the final state only comes from the thermal decoherence process instead
%of the intimal mechanical thermal occupancy as shown in Eq.~(\ref{hb}).

%%%%%%%%%%%%%%%%%%%%%%%%%%%%%%%%%%%%%%%%%%%%%%%%
%%%%%%%%%%%%%%%%%%%%%%%%%%%%%%%%%%%%%%%%%%%%%%%%
%%%%%%%%%%%%%%%%%%%%%%%%%%%%%%%%%%%%%%%%%%%%%%%%
\subsection{Comparison of intra-cavity state transfer protocols}
\label{subsec:compare}

%%%%%%%%%%%%%%%%%%%%%%%%%%%%%%%%%%%%%%%%%%%%%%%%
\begin{figure}[tbp]
\begin{center}
\includegraphics[width= 0.7 \columnwidth]{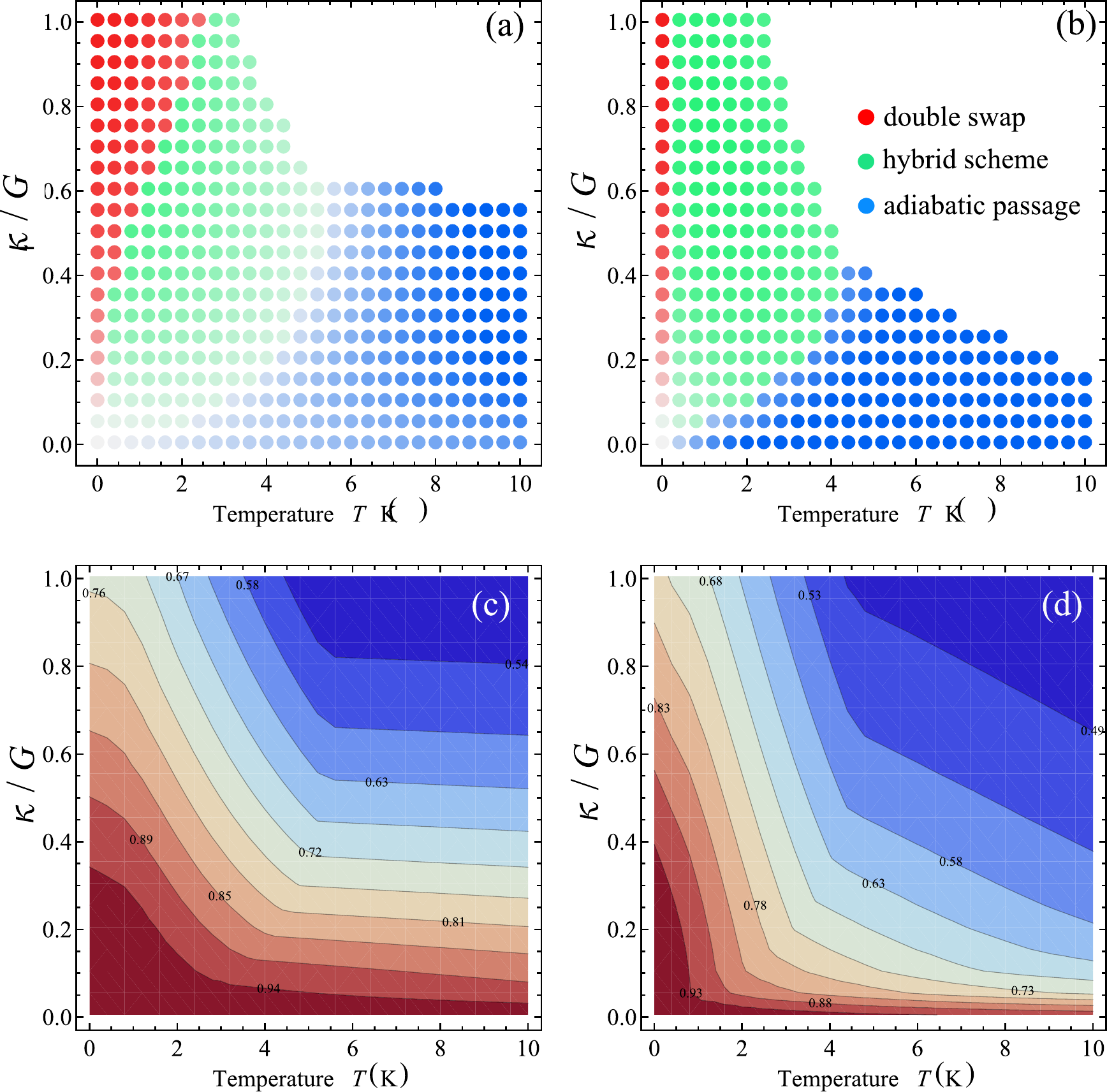}
\end{center}
\caption{(Color online) (a)  A ``phase diagram" showing the transfer protocol which yields
the highest fidelity for the transfer of a coherent state of amplitude $\alpha = 1$, as a function of environmental temperature
and cavity damping, for a fixed optomechanical coupling $G_{1,\max} = G_{2,\max} = 2
 \pi \times 2$~MHz, and for $ \gamma=2 \pi \times 100$~Hz .  Red points denote the double-swap
protocol, green points the hybrid scheme, and blue points
the adiabatic passage scheme. The saturation of each color is proportional to the
fidelity difference between the optimal scheme and the remaining two schemes.  We
only show the points where the transfer fidelity $F>0.60$.  We also take
$ \omega_M/2 \pi=10$~MHz, $\Omega_1/2 \pi=10$~GHz, and $\Omega_2/2 \pi=100$~THz.
(c) Contour plot showing fidelity of the optimal schemes, same parameters as (a).
(b) and (d):  Same as (a) and (c), except now the mechanical damping is increased to
 $ \gamma=2 \pi \times 1 $~KHz.}
% ``phase diagrams" (upper) and contour plots (lower)
%of the optimal scheme with respect to the environment temperature (in the unit of Kelvin) and the
%cavity damping (in unit of $\kappa/G$ with fixed $G$). The left column ((a) and (c)) are plots for $ \gamma=2 \pi
%\times 100$~Hz, while the right column ((b) and (d)) are plots for $ \gamma=2 \pi
%\times 1 $~KHz. In the phase diagram, the red points denote the double-swap
%protocol. The green points denote the hybrid scheme. The blue points denote
%the adiabatic passage scheme. The saturation of each color is proportional to the
%advantage (in terms of fidelity) of one scheme over the other three. Here we
%only show the points with fidelity $F>60\%$. In the contour plots ((c) and (d)), the
%fidelity of the optimal scheme (for each point) is shown and the values
%marked on the contours are the fidelity. The rest of the parameters used in these plots are $G/2
% \pi=2$~MHz, $ \omega_M/2 \pi=10$~MHz, $\Omega_1/2
% \pi=10$~GHz, and $\Omega_2/2 \pi=100$~THz.}
\label{fig:phasediagram}
\end{figure}
%%%%%%%%%%%%%%%%%%%%%%%%%%%%%%%%%%%%%%%%%%%%%%%%

Having described the three different schemes for intra-cavity state transfer sketched in Fig.~\ref{fig:pulse}, we can now
turn to compare them and discuss their relative merits.  We stress that in the absence of any cavity or mechanical dissipation, all three schemes are able to achieve perfect state transfer of an arbitrary state.  However, in the presence of dissipation, this will not be the case; moreover, the fidelity of each scheme will depend on the particulars of the dominant dissipation as well as magnitude of coupling strengths.

To allow for an easy visual assessment of this competition, we have plotted in Fig.~\ref{fig:phasediagram}(a) and (b) ``phase-diagrams" which indicate the optimal scheme for transferring a $|\alpha=1\rangle$ coherent state, as a function of both environmental temperature (taken to be the same for cavity and mechanical baths) and cavity damping rate $\kappa$. These two axes characterize the dominant dissipation mechanics:  heating from the mechanical bath, and amplitude decay due to the cavity damping.  A different color (as indicated in the figure) is used to represent each of the three schemes; for each point, the colour corresponding to the highest-fidelity scheme is plotted, with an intensity proportional to the difference of the highest fidelity between the average fidelities of all 3 schemes. For simplicity, we have use symmetric cavity parameters $G_{1,\max} = G_{2,\max} = G$ and $\kappa_1 = \kappa_2 = \kappa$.

The bottom left corner of both Fig.~\ref{fig:phasediagram}(a) and (b) corresponds to the limit where dissipative effects are minimal (i.e.~low bath temperature and low cavity damping); as expected, all three schemes have almost equal performance.  As we move away from the origin, dissipation is increased, and the near-equivalence of all three schemes in terms of fidelity is lost.
As dissipation is increased, the fidelity of all
the schemes considered will decrease, as shown in Fig.~\ref{fig:phasediagram} (c) and (d).
Note that all the schemes considered here take a time at least on the order of $1/G$ to complete, and hence require
the many-photon strong coupling condition $G > \kappa$ to achieve a good fidelity.
However, within this strong coupling regime, the
performance of each scheme varies due to their different sensitivity to
dissipation mechanisms:

\begin{itemize}
\item {The double-swap scheme (red dots in Fig.~\ref{fig:phasediagram}) is fast but it swaps the entire state into the
mechanical mode as an intermediate step. Thus it is the scheme most fragile to the effects of mechanical heating,
but the most robust against the effects of cavity damping. For an
experimental setup that is able to achieve extremely low temperatures but cannot go deep into the strong coupling regime
$G \gg \kappa$, the double swap protocol will be the most suitable protocol to
follow, as shown in the upper left region of Fig.~\ref{fig:phasediagram}.  Note from Eq.~(\ref{ds}) that the importance of amplitude decay
increases as one increases the average amplitude of the state to be transferred; this suggests the double-swap scheme
would also have an advantage for the transfer of such states. On the other hand, the importance of mechanical heating becomes even
stronger for squeezed states; we expect this to be also true for non-classical states having fine features in phase space.  Thus, the
double-swap scheme will be less advantageous for transferring such states.
}

\item {In contrast, the adiabatic scheme (blue dots in Fig.~\ref{fig:phasediagram})
is largely immune to mechanical thermal noise; this protection is due to its use
of the mechanically-dark mode. It is however easily degraded by cavity damping due to its necessarily
slow transfer time ($\gg 1 / G$). It is the best strategy when the effects of mechanical heating
completely dominate the effects of cavity decay, as shown in the lower right region of Fig.~\ref{fig:phasediagram}.
As discussed above, the heating effect becomes even more deleterious if transferring states with fine structure in phase space (such as squeezed states); the adiabatic scheme can thus be expected to be more optimal than the double-swap scheme for transferring such states.}

\item {The hybrid scheme (green dots in Fig.~\ref{fig:phasediagram}) utilizes
both the mechanically-dark mode and the coupled-cavity modes for state transfer. The resulting fast transfer time and
partial immunity to mechanical thermal noise make it the optimal scheme
when both temperature and cavity damping are equally problematic, as shown in the
intermediate region of Fig.~\ref{fig:phasediagram}. In addition, the hybrid scheme
does not require one to initially prepare the mechanical resonator in the ground state; this could also present a strong
practical advantage.
}
\end{itemize}

\section{Itinerant state transfer}

\label{section:itinerant}

The previous section was entirely devoted to three schemes for transferring an intra-cavity state between two cavities.  The state to be transferred is first prepared in one cavity, and then (via the interaction with the mechanical resonator) is transferred to the state
of the second cavity.  While this could have many useful applications (as already discussed), we have shown that all the intra-cavity transfer schemes require one to at least approach the many-photon strong-coupling condition, $G>\kappa$.  In this section, we now switch gears
and consider a state-transfer task that is possible even if one is not in this strong-coupling regime. This alternative task is the transfer of itinerant photons, i.e. converting an incoming wave-packet on cavity $1$ into an outgoing wave-packet leaving cavity $2$. Such conversion of itinerant photons is also something that could be of extreme utility, and has been discussed in several recent works~\cite{Safavi2011njp,Wang2012,Tian2012}. Experimental demonstration based on mechanical resonator and a single optical cavity has been carried out recently~\cite{Dong2012}. We expand here on the discussion in~\cite{Wang2012}, providing additional details and heuristic explanations for the phenomena which make itinerant state transfer possible in the two-cavity optomechanical system under consideration.

We start by noting that the basic physics of transferring a
narrow-bandwidth pulse is very different from that involved in transferring an intra-cavity state.
As the input state is centered around a single frequency, the transfer can essentially be viewed as a
stationary scattering process from cavity-1 input to cavity-2 output. As first noted in Ref.~\cite{Safavi2011njp}, high-fidelity
transfer thus just becomes a set of requirements on the scattering matrix characterizing our system at a single frequency.  The result is
that high-fidelity can be achieved with a more modest set of parameters.  One does not need to have many-photon
strong coupling for each cavity (i.e.~$G_i \gg \kappa$), but (as we show) the more modest requirement of a large optomechanical co-operativity: $C_i \gg 1$ for each cavity, where
\begin{equation}
    C_i \equiv \frac{G^2_i}{\kappa_i \gamma }.
    \label{eq:Coop}
\end{equation}
%
% Therefore the strong coupling condition $G\gg \kappa $ which is indispensable to the intra-cavity state transfer is
%not required. Instead, as we show below, the relevant constraint on the
%coupling strength is $G^{2}\gg \kappa \gamma $, which characterizes the
%imperfection of the scattering matrix. In this sense, the itinerant state
%transfer is less demanding than the intra-cavity state transfer. However, on
%the other hand, to transfer a narrow-bandwidth wave packet requires long
%time drives on the cavities, so experimentally one has to carefully deal
%with the induced extra heating.

In what follows, we solve for the scattering matrix describing our system, and use it to understand how and why one can achieve high-fidelity transfer of itinerant states.  In particular, we demonstrate the mechanically dark-mode introduced in Sec.~\ref{subsec:darkmode}
also plays a crucial role in itinerant state transfer, allowing it to be effective in both regime of strong and weak optomechanical coupling.  We consider both the transfer of Gaussian wavepackets, where again the simple form of Eq.~(\ref{fg}) for the transfer fidelity holds, and the transfer of arbitrary non-classical states.

\subsection{Langevin-Heisenberg equation and scattering matrix}

\label{subsec:langevin}
%
%We have discussed in Sec.~\ref{subsec:adiabatic} that the use of dark-mode
%can provide protections against thermal noise and improve the transfer
%fidelity in certain regime. We now show that the mechanically-dark mode
%discussed above plays an important role in this itinerant-photon state
%transfer, and even allows it to be highly effective in regimes of both
%strong and weak optomechanical coupling.

The itinerant state transfer can be conveniently studied using the
Heisenberg-Langevin equations of the whole system.  Working as always in an interaction
picture with respect to the main cavity drive frequencies, these take the form~\cite%
{Gardinerbook,ClerkRMP}:
\begin{eqnarray}
\dot{\hat{a}} &=&-i\omega _{M}\hat{a}-\frac{\gamma }{2}\hat{a}-i\sum G_{i}\hat{d}_{i}-\sqrt{\gamma }\hat{a}_{\text{in}}  \notag \\
\dot{\hat{d}}_{i} &=&i\Delta _{i}\hat{d}_{i}-\frac{\kappa _{i}}{2}\hat{d}_{i}-iG_{i}\hat{a}-\sqrt{\kappa _{i}}\hat{d}_{i,\text{in}}-\sqrt{\kappa
_{i}^{\prime }}\hat{d}_{i,\text{in}}^{\prime }  \label{eqs:Langevins}
\end{eqnarray}%
with  $\hat{d}_{i,\text{in}}$ representing both
input noise (taken to be white) and signals driving each cavity.  In particular, the itinerant pulse incident on cavity 1 that we wish to transfer will be described by the operator $\hat{d}_{1,{\rm in}}$.
$\kappa _{i}$ is the damping rate of cavity $i$ arising from its
coupling to the waveguide used to drive it and to extract signals from it.  In contrast, $\kappa _{i}^{\prime }$ is the
damping rate of cavity $i$ due to intrinsic losses, and $\hat{d}_{i,\text{in}}^{\prime }$ represents the corresponding noise.
Finally, $\hat{a}_{\text{in}}$ describes the thermal and quantum noise driving the mechanical resonator from its intrinsic dissipative bath.
%Strictly speaking, $\hat{a}_{in}$
%represents the mechanical intrisic noise, however we don't use $\hat{a}_{in}^{\prime }$ to represent it in order to be consistent with our previous
%paper \cite{Wang2012}.

Solving Eq.~(\ref{eqs:Langevins}) and using standard input-output relations~\cite{Gardinerbook} yield the relation between input and output fields
\begin{equation}
\hat{\vec{A}}_{\text{out}}[\omega ]=\mathbf{s}[\omega ]\hat{\vec{A}}_{
\text{in}}[\omega ]+\mathbf{s}^{\prime }\left[ \omega \right] \hat{\vec{A}}_{\text{in}}^{\prime }[\omega ]
\end{equation}%
with the vector of operators $\hat{\vec{A}}$ defined as
\begin{equation}
\hat{\vec{A}}=\left( \hat{d}_{1}[\omega ],\hat{d}_{2}[\omega ],\hat{a}
[\omega ]\right) . \label{eq:AvecDefn}
\end{equation}%
\begin{equation}
\hat{\vec{A}}^\prime=\left( \hat{d}_{1}^\prime[\omega ],\hat{d}_{2}^\prime[\omega ]\right) .
\end{equation}%
Here, $\mathbf{s}$ is the $3 \times 3[\omega_M]$ matrix describing scattering between the cavity $1$ waveguide, the cavity $2$ waveguide, and the
thermal bath responsible for the mechanical dissipation.  Note that in the general case where we have internal cavity losses, $\mathbf{s}$  will not be unitary.  In contrast, $\mathbf{s}^\prime$ is a $3 \times 2$ matrix which describes how noise from the sources of internal cavity loss can appear in the waveguides and in the mechanical bath.

Both the above matrices can be found using
Eq.~(\ref{eqs:Langevins}) and the standard input-output relations
\begin{equation}
\hat{\vec{A}}_{\mathrm{out}}[\omega ]=\hat{\vec{A}}_{\mathrm{in}}[\omega ]+\left( \sqrt{\kappa _{1}},\sqrt{\kappa _{2}},\sqrt{\gamma }\right) \cdot \hat{\vec{A}}.
\end{equation}

The elements are given by ($i=1,2$ denotes the two cavities)
\begin{eqnarray}
s_{ii}\left[ \omega \right]  &=&1-2\mu _{i}\left[ \omega \right] +C_{i}\zeta
\left[ \omega \right] \mu _{i}^{2}\left[ \omega \right] . \\
s_{12}\left[ \omega \right]  &=&s_{21}\left[ \omega \right] =\sqrt{C_{1}C_{2}
}\mu _{1}\left[ \omega \right] \mu _{2}\left[ \omega \right] \zeta \left[
\omega \right] ,\label{s12w} \\
s_{3i}\left[ \omega \right]  &=&s_{i3}\left[ \omega \right] =\frac{i}{2}
\sqrt{C_{i}}\mu _{i}\left[ \omega \right] \zeta \left[ \omega \right] , \\
s_{33}\left[ \omega \right]  &=&1-\frac{1}{4}\zeta \left[ \omega \right] ,
\end{eqnarray}
with%
\begin{equation}
\zeta \left[ \omega \right] =\left( \frac{1}{8\mu _{\mathrm{M}}\left[ \omega
\right] }+\frac{1}{2}\sum_{i}C_{i}\mu _{i}\left[ \omega \right] \right) ^{-1}
\end{equation}%
and $\mu \left[ \omega \right] $ is the susceptibility
\begin{equation}
\mu _{i}\left[ \omega \right] =\frac{\frac{\kappa _{i}}{2}}{\frac{\kappa
_{i}+\kappa _{i}^{\prime }}{2}-i\left( \omega +\Delta _{i}\right) },\ \ \mu
_{\mathrm{M}}\left[ \omega \right] =\frac{\frac{\gamma }{2}}{\frac{\gamma }{2}-i\left( \omega -\omega _{M}\right) }
\end{equation}%
and $C_{i}$ the cooperativity of cavity $i$ was defined in Eq.(\ref{eq:Coop}).
%\begin{equation}
%C_{i}=\frac{G_{i}^{2}}{\gamma \kappa _{i}}
%\end{equation}
%Notice that $\mathbf{s}[\omega ]$ is not unitary; it would only be unitary
%in the limit of no internal losses $\kappa _{i}^{\prime }=0$.

The intrinsic noise scattering matrix $\mathbf{s}^{\prime }\left[ \omega \right] $ can also be calculated in the same way
\begin{equation}
s_{ij}^{\prime }\left[ \omega \right] =s_{ij}\left[ \omega \right] \sqrt{\frac{\kappa _{j}^{^{\prime }}}{\kappa _{j}}}
\end{equation}

\begin{figure}[tbp]
\begin{center}
\includegraphics[width= 0.7 \columnwidth]{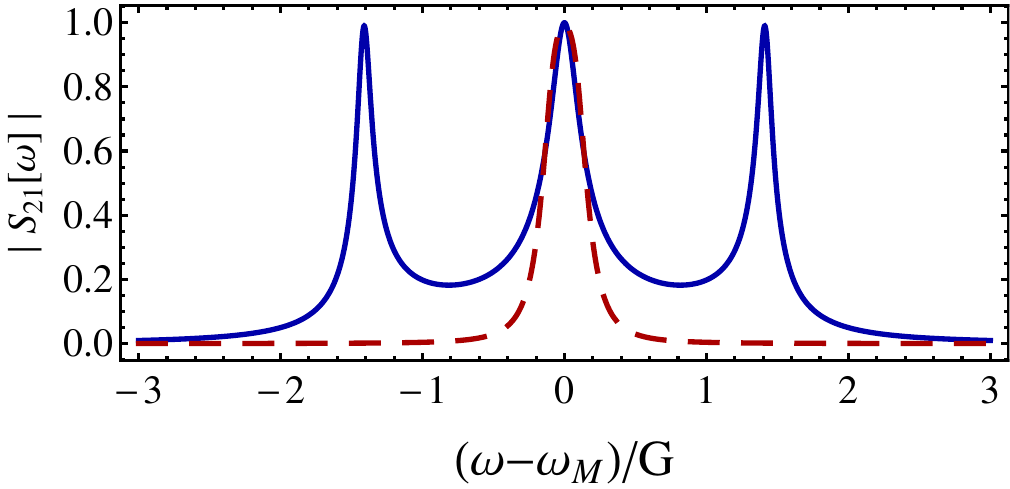}
\end{center}
\caption{(Color online) The complex modulus of the scattering matrix $|s_{21}[ \omega]|$ (Eq.~(\ref{s12w}) with no internal cavity losses, i.e.~$r_i=1$) when cavities are in resonant with the mechanics $\Delta_i=-\omega_M$;  $s_{21}[ \omega]$ describes the transmission
of a signal incident on cavity 1 to a signal leaving cavity 2 (and vice-versa). The two curves are plotted with the same mechanical damping rate $\gamma$. The solid/blue curve is obtained with $\kappa/G=0.1$ and $G/\gamma=10^3$. The red dashed curve is obtained with $\kappa/G=2$ and $G/\gamma=10^2$.}
\label{fig:scatteringmatrix}
\end{figure}

Assuming as always that large drive applied to each cavity (which yields the driven optomechanical couplings) are at the red-detuned sideband, i.e.~$\Delta _{i}=-\omega _{M}$, the configuration of the scattering
matrix is show in Fig.~\ref{fig:scatteringmatrix}. As one can see, for both
strong coupling (solid) and weak coupling (dashed), there is a transmission
window peaked around frequency $\omega _{M}$. Notice that the equations are in the rotating frame with respect to the drive frequency. Hence $\omega = \omega_M$ corresponds in the lab frame to signals at the resonant frequency of each cavity.  At this frequency the scattering elements become
\begin{eqnarray}
s_{ii}\left[ \omega _{M}\right]  &=&1-r_{i}\left( 2-\frac{8C_{i}}{\tilde{C}}
r_{i}\right) \text{, }  \label{s3i} \\
s_{12}\left[ \omega _{M}\right]  &=&s_{21}\left[ \omega _{M}\right] =\frac{8\sqrt{C_{1}C_{2}}}{\tilde{C}}r_{1}r_{2}  \label{s12} \\
s_{3i}\left[ \omega _{M}\right]  &=&s_{i3}\left[ \omega _{M}\right] =\frac{4i\sqrt{C_{i}}}{\tilde{C}}r_{i}\text{, } \\
s_{33}\left[ \omega _{M}\right]  &=&1-\frac{2}{C}
\end{eqnarray}%
with
\begin{equation}
r_{i}=\kappa _{i}/\left( \kappa _{i}+\kappa _{i}^{\prime }\right) ,
\end{equation}
and
\begin{equation}
\tilde{C}=1+4C_{1}r_{1}+4C_{2}r_{2}.
\end{equation}

In the limit of symmetric cavities (i.e.~$r_{i}=r$ and $C_1 = C_2 = C$), and high co-operativity $C \gg 1$,
the scattering matrix reduces to:
\begin{eqnarray}
s_{11}\left[ \omega _{M}\right]  &=&s_{22}\left[ \omega _{M}\right] \approx
1-r,\text{ \ }s_{33}\left[ \omega _{M}\right] =1-\frac{r}{4C} \\
s_{12}\left[ \omega _{M}\right]  &=&s_{21}\left[ \omega _{M}\right] \approx r
\\
s_{31}\left[ \omega _{M}\right]  &=&s_{13}\left[ \omega _{M}\right] =s_{23}\left[ \omega _{M}\right] =s_{32}\left[ \omega _{M}\right] =\frac{i}{2\sqrt{C}}\text{. }  \label{s3}
\end{eqnarray}
We thus see that transmission at the cavity frequency becomes perfect in the limit of no internal loss (i.e. $r \ra 1$); in this limit, the
above scattering matrix reduces to the results of~\cite{Wang2012}. In our following discussions of itinerant transfer, unless specified, we take this limit for simplicity.

High fidelity transfer from $\hat{d}_{1,\mathrm{in}}$ to $\hat{d}_{2,\mathrm{out}}$ requires that over the input signal bandwidth, the transmission
coefficient $|s_{21}[\omega ]|^{2}\sim 1$, as well as that $|s_{23}[\omega
]|^{2}\sim 0$ (i.e.~negligible transmission of mechanical noise).  Of course, if there is no internal loss, the scattering matrix $\mathbf{s}$ is unitary, and hence the former condition implies the latter condition.  We see from the above results that
in the limit of low internal cavity losses, these ideal scattering conditions are satisfied when the input signal is centered at $\omega _{M}$ and the system parameters $C_{i}=C\gg 1$. In the following subsection, we give physical
explanations of these perfect transmission conditions.

\subsection{Dark mode transfer due to Optomechanical EIT mechanism}

\label{subsec:eit}

While the basic results for itinerant photon transfer follow immediately from the elementary calculation of the scattering
matrix presented in the previous subsection, these calculations do not necessarily yield a good intuitive feel
for why such ideal scattering conditions can be realized.  Such a heuristic understanding is the goal of this subsection; in particular, we show that the mechanically-dark mode plays an essential role in achieving a high
fidelity itinerant state transfer.

To have protection against mechanical dissipation, one would ideally like
the input state incident on cavity 1 to only excite the dark mode. Without
dissipation, the dark mode $\hat{c}_{\rm{dk} }$ defined in Eq.~(\ref{dkm}) is
energetically separated from the mixed modes $\hat{c}_{\pm }$ defined in Eq.
(\ref{cpm}). If the input signal is closely centered at frequency $\omega
_{M}$ in the displaced frame, only the dark mode is excited and hence
protection against mechanical thermal noise is achieved.

Including dissipation (and the consequent lifetime broadening), the input
signal incident on cavity 1 will also excite the bright cavity mode $\hat{c}
_{\text{br}}$ (see Eq. (\ref{brm})) as well as the mechanical mode $\hat{a}$. This is not a issue in the strong coupling regime, as the transmission
channel of the dark mode (central peak of solid line in Fig.~\ref
{fig:scatteringmatrix}) is well separated from the bright modes transmission
(two side peaks of the solid line in Fig.~\ref{fig:scatteringmatrix}).
However, a problem arises in the weak coupling limit $\kappa >G$: the
broadening of each mode is larger than their separation, and thus they are not
resolvable from one another (see red dashed line in Fig.~\ref%
{fig:scatteringmatrix}).  Thus, for weak coupling, one cannot selectively drive only the dark mode
by simply tuning the frequency of the input signal to be $\omega_M$.

Fortunately, as we show below, in the weak-coupling regime, the unwanted
excitation is irrelevant as long as the cooperativity of each cavity $C_{i} \gg 1$. In this limit, the bright mode
amplitude $\langle \hat{c}_{\text{br}}\rangle $ arising from the signal
incident of cavity 1 is a factor $\sim 1/C$ smaller than the dark mode
amplitude, due to a destructive interference akin to the optomechanical
analogue of electromagnetic-induced transparency~\cite%
{Agarwal2010,Weis2010,Safavi2011}.

In order to understand this, it is useful to introduce a ``mixed" $3 \times 3$
susceptibility matrix $\boldsymbol{\chi }[\omega ]$ which describes
how the input modes incident on each resonator drive the non-local system modes $\hat{\vec{C}}[\omega ]=\{\hat{c}_{\mathrm{dk}}[\omega ],\hat{c}_{\mathrm{br}}[\omega ],\hat{a}[\omega ]\}$:
\begin{equation}
\hat{\vec{C}}[\omega ]=\boldsymbol{\chi }[\omega ]\mathbf{k}\hat{\vec{A}}_{%
\mathrm{in}}[\omega ],
\end{equation}%
where $\mathbf{k}=\mathrm{diag}\{\sqrt{\kappa _{1}},\sqrt{\kappa _{2}},\sqrt{\gamma }\}$ is a diagonal matrix,
and the vector $\hat{\vec{A}}_{\rm in}$ is defined in Eq.~(\ref{eq:AvecDefn}).

There are two main inputs driving this
system: the signal from cavity 1 and the thermal noise from the mechanics.
Hence the elements of particular interest are $\chi _{j,1}$ and $\chi _{j,a}$ ($j=\left\{ \mathrm{dk},\mathrm{br},a\right\} $). The matrix $\boldsymbol{\chi }[\omega ]$ can be found directly from Eq.~(\ref{eqs:Langevins}). While
the full expression is rather lengthy, at frequency $\omega =\omega _{M}$,
it can be written as (assuming no internal cavity losses)
\begin{equation}
\boldsymbol{\chi }[\omega ]=\frac{2}{4C_{1}+4C_{2}+1}\left(
\begin{array}{ccc}
-\tilde{G}_{2}M_{2} & -\tilde{G}_{1}M_{1} & -2i\frac{\kappa _{1}-\kappa _{2}
}{\sqrt{G_{1}^{2}+G_{2}^{2}}}\sqrt{C_{1}C_{2}} \\
\frac{\tilde{G}_{1}}{\kappa _{1}} & \frac{\tilde{G}_{2}}{\kappa _{2}} & -2i\frac{C_{1}+C_{2}}{\sqrt{G_{1}^{2}+G_{2}^{2}}} \\
-2i\frac{G_{1}}{\gamma \kappa _{1}} & -2i\frac{G_{2}}{\gamma \kappa _{2}} &
\frac{1}{\gamma }%
\end{array}%
\right)
\end{equation}%
with%
\begin{equation}
\tilde{G}_{i}\equiv \frac{G_{i}}{\sqrt{G_{1}^{2}+G_{2}^{2}}},M_{i}=4\left(
\frac{C_{1}}{\kappa _{2}}+\frac{C_{2}}{\kappa _{1}}\right) +\frac{1}{\kappa
_{i}}
\end{equation}

Consider first the excitation of the \textquotedblleft bright" mode
$\hat{c}_{\mathrm{br}}$ relative to the dark mode $\hat{c}_{\mathrm{dk}}$ by
a signal incident on cavity 1 at frequency $\omega =\omega _{M}$. In the
ideal limit where $C_{1}=C_{2}\equiv C\gg 1$, we find:
\begin{equation}
\frac{\chi _{\mathrm{br},1}\left[ \omega _{M}\right] }{\chi _{\mathrm{dk},1}\left[ \omega _{M}\right] }\approx -\frac{1}{4C}\frac{\sqrt{\kappa
_{1}\kappa _{2}}}{\left( \kappa _{1}+\kappa _{2}\right) }
\end{equation}%
Thus, the relative amplitude of the bright mode is suppressed by a large
factor $C$ compared to the dark mode, {\it even if} one does not have $G \gg \kappa$ and hence a strong
spectral separation of these modes. Physically, this is due to a EIT-like
destructive interference.  In the absence of any optomechanical coupling (i.e.~$G=0$),  an input signal incident on cavity $1$
would excite both the dark mode and the bright mode equally.  However, when $G \neq 0$, the coupling
between the bright mode and mechanics strongly modifies its susceptibility (this is the essence of optomechanical EIT).
Heuristically, an excitation
on the bright mode will be swapped multiple times between the bright mode and the
mechanical mode, resulting in destructive interference. In
case of two symmetrical cavities, the modified susceptibility due to these
``multiple swaps" can be represented by a Dyson series:
\begin{equation}
\chi _{\mathrm{br},1}=\chi _{\mathrm{br},1}^{\left( 0\right) }\left(
1+G^{2}\chi _{\mathrm{br}}^{\left( 0\right) }\chi _{a}^{\left( 0\right)
}+\left( G^{2}\chi _{\mathrm{br}}^{\left( 0\right) }\chi _{a}^{\left(
0\right) }\right) ^{2}+...\right) =\frac{\chi _{\mathrm{br},1}^{\left(
0\right) }}{1-G^{2}\chi _{\mathrm{br}}^{\left( 0\right) }\chi _{a}^{\left( 0\right)
}}=\frac{\chi _{\mathrm{br},1}^{\left( 0\right) }}{1+C}  \label{sp}
\end{equation}%
where have focus on the ideal frequency $\omega = \omega_M$.
Here, $\chi _{\mathrm{br},1}^{\left( 0\right) }\approx -\sqrt{2/\kappa _{1}}$
is the susceptibility of the bright (dark) mode to the input of cavity 1
when $G\approx 0$, and $\chi _{a}^{\left( 0\right) }=-i/\gamma $ is the
susceptibility of the mechanical mode, $\chi _{\mathrm{br}}^{\left( 0\right)
}=-i/\kappa $ is the susceptibility of the bright mode to the mechanical
mode in resonance. As we see, the coupling of the dark mode to the
mechanical modes results in the suppression factor $1/\left( 1+C\right) $ in
Eq.~(\ref{sp}). This phenomenon identical to optomechanical
EIT, where the coupling of the optical mode with the mechanical mode prevents the cavity
from being excited by an incident signal.  The cancellation is valid as
long as the the cooperativity is large, and independent of the strong
coupling condition.

On the other hand, the input signal on cavity $1$ will also excite the
mechanical mode $\hat{a}$.  Again using our mixed-susceptibility, one finds
\begin{equation}
\frac{\chi _{a,1}\left[ \omega _{M}\right] }{\chi _{\mathrm{dk},1}\left[
\omega _{M}\right] }\approx \frac{i}{2\sqrt{\gamma C}}\sqrt{\frac{\kappa
_{1}\kappa _{2}}{\kappa _{1}+\kappa _{2}}}\propto \frac{\kappa }{G}
\end{equation}%
Thus, if one is in the regime of weak coupling $G_{i}/\kappa _{i}<1$, the
excitation of the mechanical mode can be appreciable. However, due to the
relatively weak coupling $\gamma $ between the mechanical resonator and its
dissipative bath, the overall contribution from this excitation of $\hat{a}$
to $\hat{a}_{\mathrm{out}}$ scales as $\sqrt{\gamma }\hat{a}$; this follows
from the input-output relation $\hat{a}_{\mathrm{out}}=\hat{a}_{\mathrm{in}}+\sqrt{\gamma }\hat{a}$. The net result is that only a small amount of the
input signal is lost to the mechanical bath: the scattering matrix element $s_{31}$ describing this process is small as $1/\sqrt{C}$ (c.f.~Eq.~(\ref{s3})). Therefore, the transfer of the input signal thus occurs almost entirely
via the dark mode in this limit.

Good fidelity also requires that the dark mode, once excited by the input
state, only leaks out via cavity 2, ensuring $|s_{21}[\omega _{M}]|\approx 1$. In another word, the input signal has little leakage through the output of
cavity 1. This requires a destructive interference between the promptly
reflected input signal and the wave leaving the dark mode via cavity 1. That
is, the signal reflected directly out of cavity 1 and the signal scattered
into cavity 1 output cancels. For $C_{i}\gg 1$, this interference
cancellation results in the simple impedance matching condition~\cite{Safavi2011njp,Wang2012} $C_{i}\equiv C$, i.e.:
\begin{equation}
G_{1}^{2}/\kappa _{1}=G_{2}^{2}/\kappa _{2}  \label{con}
\end{equation}

We have thus shown how the incident signal primarily excites the mechanically-dark mode of the two-cavity optomechanical system, and how the dark mode then leaks out to cavity 2.  It remains to show that we also have a suppression of noise originating from the mechanical bath in the output of cavity $2$.  Eq.~(\ref{s3}) shows that for $C\gg 1$, $s_{23}\left[ \omega _{M}\right] $ is suppressed by a small factor $1/\sqrt{C}$; this directly
results in the small value of the output heating parameter $\bar{n}_{\text{h}}$ discussed in the main text (see Eq. (9)). This suppression is most easily
understood in the case where $\kappa _{1}=\kappa _{2}$. In this case, the
dark mode is completely decoupled from the mechanical mode. The bright mode
is however driven by the mechanical noise, and thus provides a route for
mechanical noise to corrupt the output from cavity 2. This process yields a
contribution $\sim \sqrt{\kappa }\cdot \chi _{\mathrm{br},a}\left[ \omega
_{M}\right] \cdot \sqrt{\gamma }$ to $s_{23}\left[ \omega _{M}\right] $. The
relevant susceptibility $\chi _{\mathrm{br},a}\left[ \omega _{M}\right] \sim
1/G$, yielding the result $s_{23}\left[ \omega _{M}\right] \sim 1/\sqrt{C}$.

The situation is only slightly more complicated when $\kappa _{1}\neq \kappa
_{2}$. In this case, the cavity decay terms in Eq.~(\ref{eqs:Langevins}) can
effectively mix the bright and the dark cavity modes.
As a result, mechanical noise can first excite the bright mode, then be
mixed into dark mode, then find its way into the cavity 2 output field. This
additional process is not problematic, as it also scales the same way with $C $ as the $\kappa _{1}=\kappa _{2}$ process. This follows from the fact
that $\chi _{\mathrm{dk},a}[\omega _{M}]\sim \left( (\kappa _{1}-\kappa
_{2})/\tilde{\kappa}\right) \chi _{\mathrm{br},a}[\omega _{M}]$, where $\tilde{\kappa}=2\kappa _{1}\kappa _{2}/(\kappa _{1}+\kappa _{2})$
(i.e.~excitation of the dark mode by mechanical noise involves first
exciting the bright mode). The net result is Eq.~(\ref{s3}): $s_{23}[\omega
_{M}]$ scales as $1/\sqrt{C}$ irrespective of $\kappa _{1}/\kappa _{2}$.

\subsection{Transfer fidelity for Gaussian states}

\label{subsec:fidelity}

To quantify the transfer fidelity of the itinerant photon transfer scheme, we consider a Gaussian input state in a
temporal mode (see, e.g.,~Ref.~\cite{ClerkRMP}) defined by
\begin{equation}
\hat{D}_{1,\text{in}}=\left( 2\pi \right) ^{-1/2}\int d\omega f\left[ \omega \right]
\hat{d}_{1,\text{in}}\left[ \omega \right],
\end{equation}%
where $f[\omega ]$ describes a wave packet
incident on cavity 1 which is localized in both frequency and time; $\int
d\omega |f[\omega ]|^{2}=1$ to ensure that $\hat{D}_{1,\text{in}}$ is a canonical
bosonic annihilation operator. $\hat{d}_{1,\text{in}}$ corresponds to an input mode with a fixed frequency, but completely delocalized in time. The fidelity of transferring this itinerant Gaussian state takes the same form as Eq.~(\ref{fg}). In case of an
itinerant coherent state input $|\psi _{\mathrm{in}}\rangle \propto \exp
\left( \alpha \hat{D}_{1,\text{in}}^{\dag }\right) |0\rangle $, the heating $\bar{n}_{\mathrm{h}}$ and the damping $\lambda $ can be explicitly written as (for
details of the input temporal mode, see \ref{app:itinerant})
\begin{eqnarray}
\bar{n}_{\mathrm{h}} &=&\sum_{i=1,2,M}\int d\omega \left\vert f\left[ \omega %
\right] s_{2i}\left[ \omega \right] \right\vert ^{2}N_{i}~~~  \notag \\
\lambda &=&|\alpha |\min_{\tau }\left( 1-\left\vert \int \,d\omega
e^{-i\omega \tau }s_{21}\left[ \omega \right] \left\vert f\left[ \omega %
\right] \right\vert ^{2}\right\vert \right) ~~~~~  \label{nl}
\end{eqnarray}%
We have optimized the final state $\hrho _\text{f}$ in Eq.~(\ref{eq:FidelityDefn}) over a
time-translation $\tau $, so that if the output pulse is simply a
time-delayed copy of the input pulse, $F=1$.

\begin{figure}[tbp]
\begin{center}
\includegraphics[width= 0.5 \columnwidth]{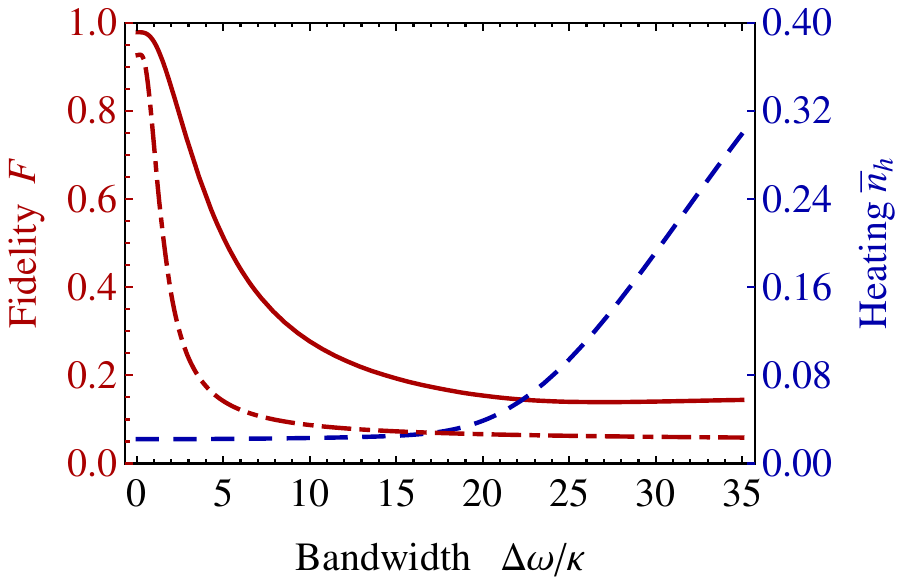}
\end{center}
\par
\vspace{-0.5cm}
\caption{ Fidelity $F$ (red) and heating $\bar{n}_\text{h}$ (blue) versus input
bandwidth for transferring $| \alpha=  \sqrt{3}\rangle$
coherent state. The red dash-dot line denotes the fidelity in the weak
coupling case, where $G/2 \pi=0.1$~MHz, $ \kappa=2 \pi \times 0.5$~KHz, and $\gamma=2\pi \times 10$~Hz. The red solid line and the blue dashed line are both for strong
coupling case where $G/2\pi=0.6$~MHz, $\kappa=2\pi\times 50$~KHz and $\gamma=2 \pi \times 1$~KHz. The temperature is $T=2$~K and the cavity frequencies are $\Omega_1/2\pi=10$~GHz, and $\Omega_2/2\pi=100$~THz. }
\label{fig:bandwidth}
\end{figure}

Taking our input mode $|f[\omega ]|^{2}$ to have mean frequency $\omega _{M}$
and a Gaussian profile with variance $\Delta \omega ^{2}$, and assuming $%
C_{1}=C_{2}=C\gg 1$, we find to leading order in $\Delta \omega $:
\begin{eqnarray}
\bar{n}_{\mathrm{h}} &\approx &\frac{N_{M}}{4C}~\left( 1+\left( \frac{\Delta
\omega }{G}\right) ^{2}\left( 1-\frac{\kappa ^{2}}{16G^{2}}\right) \right) ~
\label{eq:itinerantneff} \\
\lambda &\approx &|\alpha |\left[ \frac{1}{8C}+\left( \frac{2\Delta \omega }{\kappa }\right) ^{2}\left( 1+\left( \frac{\kappa ^{2}}{8G^{2}}\right)
^{2}\right) \right]  \label{eq:itinerantampdecay}
\end{eqnarray}
Good fidelity requires a high cooperativity $C\gg |\alpha |,N_{M}$. In the
weak-coupling regime $G<\kappa $, one also needs $\sqrt{|\alpha |}\Delta
\omega \ll (G^{2}/\kappa )$, which reflects the width of the $s_{21}[\omega
] $ transmission resonance. In the opposite regime $G\gg \kappa $, one needs
$\Delta \omega \leq \kappa /\sqrt{|\alpha |}$ as shown in Fig.~\ref{fig:bandwidth}.

In the limit $\Delta \omega \rightarrow 0$, Eq.~(\ref{eq:itinerantneff},\ref{eq:itinerantampdecay}) reduces to
\begin{equation}
\bar{n}_{\mathrm{h}}\approx \frac{N_{M}}{4C},~~\lambda \approx \frac{|\alpha
|}{8C}
\end{equation}
which shows the fidelity of itinerant state transfer depends solely on the
cooperativity in case of zero-bandwidth. Further, we see that in comparison
against the double-swap scheme, the mechanical-heating effect described by $\bar{n}_{\mathrm{h}}$ is reduced by a factor $\kappa /G$. The expression of $\bar{n}_{\mathrm{h}}$ is the usual  mechanical temperature expression of the cavity cooling in the weak coupling regime~\cite{Marquardt2007,Wilson2007};
unlike cavity-cooling, it describes $\bar{n}_{\mathrm{h}}$ in both weak and
strong coupling regimes.

\subsection{Transfer of non-classical itinerant states}

\label{subsec:nongaussian}

Given the advantages of the itinerant transfer scheme, it is also
interesting to consider how well it is able to transfer non-classical
states. While in general it is difficult to obtain analytic expressions for
the evolution of non-Gaussian states, we show that here, one can obtain
useful and reliable analytic approximations.

We again consider an input mode in a given temporal mode $\hat{D}_{1,\text{in}}$; we
take this mode to be centered on $\omega _{M}$, and for simplicity, to have
a vanishingly small bandwidth $\Delta \omega $. Suppose now the input state
incident on cavity 1 is
\begin{equation}
\hrho _{in,1}=\sum_{mn}c_{n}^{\ast }c_{m}\left\vert m\right\rangle _{1,\text{\textrm{in}}}\left\langle n\right\vert
\end{equation}%
where $|n\rangle _{1,\text{\textrm{in}}}$ is a Fock state of this mode $|n\rangle _{1,\text{\textrm{in}}}\propto \left( \hat{D}_{1,\text{in}}^{\dag }\right)
^{n}|0\rangle $. We also take the noise driving both cavities to be
zero-temperature ($N_{1}=N_{2}=0$), but allow the mechanical resonator to be
driven by thermal noise. Letting $p_{\mathrm{th}}(q,N_{m})$ be the
probability of having $q$ thermal quanta incident on the mechanical
resonator, the fidelity can be decomposed as (see \ref{app:nongaussian} for
more details)%
\begin{equation}
F=\sum_{q,r=0}^{\infty }p_{\mathrm{th}}(q,N_{m})\sum_{mn}
\sum_{d=-n}^{q-r}c_{m+d}^{\ast }c_{n+d}c_{n}^{\ast
}c_{m}f_{m,0,q}^{r,m+d,q-r-d}\left( f_{n,0,q}^{r,n+d,q-r-d}\right) ^{\ast },
\label{fs}
\end{equation}%
with
\begin{eqnarray}
f_{n,0,q}^{r,n+d,q-r-d} &=&\sqrt{\frac{r!n!q!\left( n+d\right) !}{\left(
q-r-d\right) !}}\left( s_{21}\right) ^{n}\left( s_{33}\right) ^{q}\left(
\frac{s_{13}}{s_{33}}\right) ^{r}\left( \frac{s_{23}}{s_{33}}\right) ^{d}
\notag \\
&&\sum_{j=0}^{\min \left[ n,r\right] }\frac{_{2}F_{1}\left( j-n,d+r-q;1+d+j;%
\frac{s_{31}s_{23}}{s_{33}s_{21}}\right) }{j!\left( j+d\right) !\left(
n-j\right) !\left( r-j\right) !}\left( \frac{s_{11}s_{23}}{s_{21}s_{13}}%
\right) ^{k},
\end{eqnarray}%
where $_{2}F_{1}$ is the hyper-geometric function and $\mathbf{s}\equiv
\mathbf{s}\left[ \omega _{M}\right] $. $f_{n,0,q}^{r,n+d,q-r-d}$ is the
amplitude for an input state $\left\vert n,0,q\right\rangle _{\text{in}}$
scattering into an output $\left\vert r,n+d,q-r-d\right\rangle _{\text{out}}$. $\left\vert n,0,q\right\rangle $ denotes the state with $n$ photon in
cavity 1, $0$ photon in cavity 2, and $q$ phonon in the mechanics. Note that
the scattering matrix conserves the total excitation, therefore the
amplitudes between states with different excitation numbers vanish.

In the case of a Fock state input $|n\rangle $, $c_{m}=\delta _{m,n}$, the
above result is reduced to%
\begin{equation}
F=\sum_{r=0}^{\infty }P(r,n)=\sum_{r=0}^{\infty }\sum_{q=0}^{\infty }p_{\mathrm{th}}(q,N_{m})\left\vert f_{q}^{\left( r,n\right) }\right\vert ^{2}
\label{fock}
\end{equation}%
where $P\left( r,n\right) $ is the probability of having $r$ outgoing
photons leaving cavity 1 and $n$ photons leaving cavity 2, and
\begin{eqnarray}
f_{q}^{\left( r,n\right) } &=&\sqrt{\left(
\begin{array}{c}
q \\
r%
\end{array}%
\right) }\left( s_{21}\right) ^{n}\left( s_{33}\right)
^{q}\sum_{j=0}^{r}\left( \frac{s_{13}}{s_{33}}\right) ^{r}\,\left(
\begin{array}{c}
n \\
j%
\end{array}%
\right) \left(
\begin{array}{c}
r \\
j%
\end{array}%
\right)  \notag \\
&&\left( \frac{s_{11}s_{23}}{s_{21}s_{13}}\right) ^{j}{}_{2}F_{1}\left(
j-n,r-q;1+j;\frac{s_{31}s_{23}}{s_{21}s_{33}}\right)
\end{eqnarray}

Note that in the regime of optimal state transfer $C_{1}=C_{2}\equiv C\gg 1$, the probability of having photons leave cavity 1 is small: the dark state
effectively prevents mechanical photons from contributing, and Eq.~(\ref{con}) ensures minimal reflection of signal photons. One can thus get a good
approximation by simply retaining the $r=0$ and $r=1$ term in Eq.~(\ref{fock}): $F$ is approximately the probability of obtaining $n$ photons in the
cavity 2 output mode and at most one photon leaving cavity 1. This is a rigorous lower bound on the exact fidelity, and is exact to order $1/C$.

\begin{figure}[tbp]
\begin{center}
\includegraphics[width= 0.5 \columnwidth]{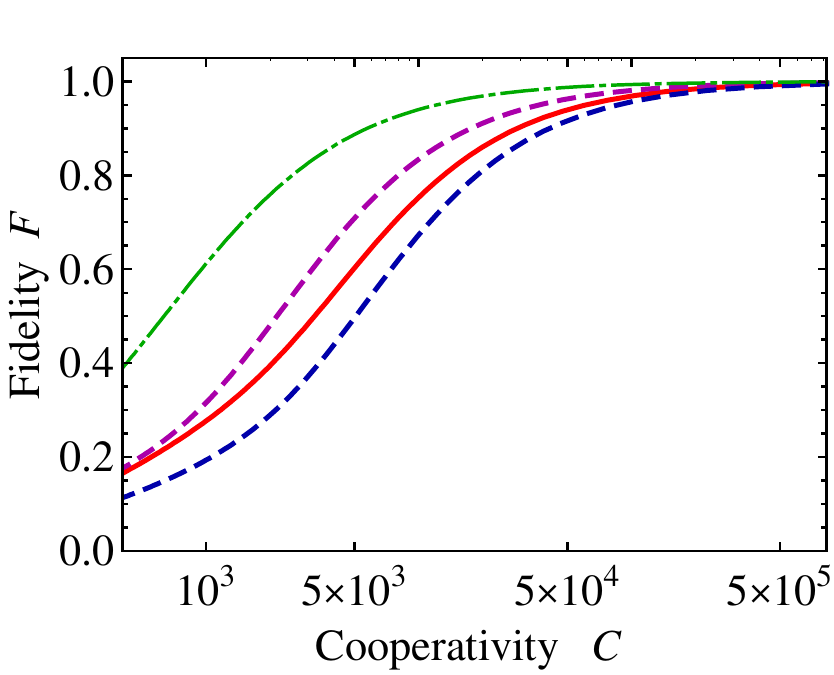}
\end{center}
\par
\vspace{-0.5cm}
\caption{ Fidelity versus cooperativity $C = G^2 /\kappa\gamma$ in the narrow-bandwidth limit. The input states are $|\alpha%
=  \sqrt{3} \rangle$ (green, dash-dot), $|n=3\rangle$ Fock state
(blue, dashed), $|n=1\rangle$ Fock state
(magenta, dashed) and $(|1\rangle+|3\rangle)/ \sqrt{2}$ (red, solid).
The coupling strength is $G/2 \pi=0.6$~MHz. Unless specified, the
parameters are the same as the right column of Fig.~ \ref%
{fig:phasediagram}. }
\label{fig:nongaussian}
\end{figure}

In the limit $C\gg 1$, one finds that to the leading order in $1/C$ the fidelity
for transferring the $n$-photon itinerant Fock state is $F\simeq 1-\left[
N_{M}\left( 3+2n\right) +n\right] /4C$. For $N_{M}\gg 1$, the condition for
a near-unity fidelity is thus $C\gg N_{M}n$; for a large-$n$ Fock state,
this is more stringent than the condition for having a large fidelity
transfer of a coherent state with $|\alpha |\sim \sqrt{n}$ (c.f.~Eqs.~(\ref{eq:itinerantneff}),(\ref{eq:itinerantampdecay})).

The transfer fidelity of different non-Gaussian states together with a
coherent state (for realistic parameters) are shown in Fig.~\ref{fig:nongaussian}. The figure shows high fidelity transfer of non-Gaussian
states is possible if the cooperativity is sufficiently large. For the same
cooperativity, the fidelity of transferring a Fock state (the blue dashed
line) is lower than that of transferring a coherent state of the same
average energy (the green dash-dot line). The difficulty probably arises
from the non-trivial quantum nature and complicated phase space structure of
the Fock state.

\section{Conclusion}

\label{conclusion}

In this paper, we have studied quantum state transfer from an optical cavity to a microwave cavity via a jointly coupled mechanical resonator. We have introduced the notion of a mechanically-dark mode which is insensitive to the mechanical thermal noise. We have proposed different approaches to utilize the dark mode for state transfer and compared their advantages and disadvantages, in particular:

1) For the transfer of an intra-cavity state, we find that a hybrid scheme which uses both the dark mode and the bright modes can achieve high transfer fidelity, even without pre-cooling the mechanical resonator to the ground state. We also provide a ``phase diagram'' to help experimentalists identify the best transfer scheme for their particular system.

2) In the transfer of itinerant photon states, the dark mode can be used through a mechanism similar to the optomechanical electromagnetic transparency (EIT). Both Gaussian and non-Gaussian states can be transferred with high fidelity if the cooperativity is sufficiently high.

In the short term, experiments are racing towards the transfer of coherent states and wave packets; we also expect the improvement due to dark modes can be demonstrated. Whereas in the long term, hopefully non-classical state transfer can be realized experimentally. This will open up new possibilities in hybrid quantum information network.

\section{Acknowledgment}

\label{acknowledgment}

We thank S. Chesi, K. Lehnert, O. Painter, C. Regal and A. Safavi-Naeini for
useful discussions. This work was supported by the DARPA ORCHID program
under a grant from the AFOSR.

\newpage

\appendix

\section{Gaussian state fidelity}

\label{app:gaussianfidelity}

Since the state to be transferred is mostly a pure state, we consider the
case where at least one of the states $\hat{\rho}_{\mathrm{i}},\hat{\rho}_{%
\mathrm{f}}$ is a pure state. Then the general definition of the transfer
fidelity in Eq.~(\ref{eq:FidelityDefn}) reduces to the simple overlap $F=%
\mathrm{Tr}\left( \hat{\rho}_{\mathrm{i}}\hat{\rho}_{\mathrm{f}}\right) $.
In terms of the Wigner representation $W_{\mathrm{i}}\left( \vec{\xi}\right)
$ \ and $W_{\mathrm{f}}\left( \vec{\xi}\right) $ of each density matrix:
\begin{equation}
F=\mathrm{Tr}\left( \hat{\rho}_{\mathrm{i}}\hat{\rho}_{\mathrm{f}}\right)
=\pi \int_{-\infty }^{+\infty }d\vec{\xi}\,W_{\mathrm{i}}\left( \vec{\xi}%
\right) W_{\mathrm{f}}\left( \vec{\xi}\right)  \label{fw}
\end{equation}%
Here, $\vec{\xi}$ is the vector formed by the quadratures of the $N$ modes
of our system ($N=1$ here): $\vec{\xi}=\left\{ x_{1},p_{1},x,p_{2},\cdots
,x_{N},p_{N}\right\} $ and $\hat{\vec{\xi}}=\left\{ \hat{x}_{1},\hat{p}_{1},%
\hat{x}_{2},\hat{p}_{2},\cdots ,\hat{x}_{N},\hat{p}_{N}\right\} $. We define
the quadratures of each mode as $\hat{x}_{i}=\left( \hat{d}_{i}+\hat{d}%
_{i}^{\dag }\right) /2$ and $\hat{p}_{i}=-i\left( \hat{d}_{i}-\hat{d}%
_{i}^{\dag }\right) /2$, where $\hat{d}_{i}$ is the canonical annihilation
operator of the mode.

As our system is described by a quadratic Hamiltonian, an initial Gaussian
state will remain Gaussian at all times. The Wigner function of a Gaussian
state can be written in general in terms of the means and covariances of the
mode quadratures:
\begin{equation}
W\left( \vec{\xi}\right) =\frac{1}{\left( 2\pi \right) ^{N}\sqrt{\det
\mathbf{V}}}\exp \left\{ -\frac{1}{2}\left( \vec{\xi}-\left\langle \hat{\vec{%
\xi}}\right\rangle \right) \cdot \mathbf{V}^{-1}\left( \vec{\xi}%
-\left\langle \hat{\vec{\xi}}\right\rangle \right) \right\}  \label{wig}
\end{equation}%
where $\mathbf{V}$ is the symmetrized covariance matrix, defined via $%
V_{jj^{\prime }}=\frac{1}{2}\left\langle \Delta \hat{\xi}_{j}\Delta \hat{\xi}%
_{j^{\prime }}+\Delta \hat{\xi}_{j^{\prime }}\Delta \hat{\xi}%
_{j}\right\rangle $, with $\Delta \hat{\xi}_{j}=\hat{\xi}_{j}-\langle \hat{%
\xi}_{j}\rangle $.

Plugging Eq. (\ref{wig}) into Eq. (\ref{fw}), we obtain%
\begin{equation}
F = \frac{\pi }{\left( 2\pi \right) ^{N}\sqrt{\det \left( \mathbf{V}_{%
\mathrm{i}} +\mathbf{V}_{\mathrm{f}}\right) }} \exp \left[ -\frac{1}{2} %
\Bigg( \left\langle \hat{\vec{\xi}}_{\mathrm{i}}\right\rangle -\left\langle
\hat{\vec{\xi}}_{\mathrm{f}}\right\rangle \Bigg) \cdot \frac{1}{ \mathbf{V}_{%
\mathrm{i}}+\mathbf{V}_{\mathrm{f}}} \Bigg( \left\langle \hat{\vec{\xi}}_{%
\mathrm{i}}\right\rangle -\left\langle \hat{\vec{\xi}}_{\mathrm{f}%
}\right\rangle \Bigg) \right]  \label{eq:FullFidelity}
\end{equation}%
If we now define
\begin{eqnarray}
\bar{n}_{\mathrm{h}} &=&2\sqrt{\det \left( \mathbf{V}_{\mathrm{i}}+\mathbf{V}%
_{\mathrm{f}}\right) }-1  \label{nb} \\
\lambda ^{2} &=& \Bigg( \left\langle \hat{\vec{\xi}}_{\mathrm{i}%
}\right\rangle -\left\langle \hat{\vec{\xi}}_{\mathrm{f}}\right\rangle %
\Bigg) \cdot \frac{\sqrt{\det \left( \mathbf{V}_{\mathrm{i}}+\mathbf{V}_{%
\mathrm{f}}\right) }}{\mathbf{V}_{\mathrm{i}}+\mathbf{V}_{\mathrm{f}}} %
\Bigg( \left\langle \hat{\vec{\xi}}_{\mathrm{i}}\right\rangle -\left\langle
\hat{\vec{\xi}}_{\mathrm{f}}\right\rangle \Bigg) ,  \label{nlam}
\end{eqnarray}%
then Eq.~(\ref{eq:FullFidelity}) becomes the simple expression for the
fidelity given in the main text, Eq.~(\ref{fg}). %we get Eq.~(%
%\ref{fg}) in the main text
%\begin{equation}
%F=\frac{1}{1+\bar{n}_{\mathrm{h}}}\exp \left( -\frac{\lambda ^{2}}{1+\bar{n}%
%_{\mathrm{h}}}\right)   \label{f}
%\end{equation}%
Note that throughout the paper, we optimize the fidelity over simple
rotations in phase space (so that if $\hat{\rho}_{\mathrm{f}}$ is a rotated
version of $\hat{\rho}_{\mathrm{i}}$, $F=1$).

\section{Analytical solutions for intra-cavity state transfer}

\label{app:intracavity}

For Gaussian states, we can rewrite the effective master equation in a
general form characteristic to bilinear Hamiltonians
\begin{equation}
\dot{\hat{\rho}}\left( t\right) =-i\left[ \hat{\vec{\xi}} \cdot \mathbf{H}
\hat{\vec{\xi}},\hat{\rho}\left( t\right) \right] + \sum_{k}\frac{\gamma _{k}%
}{2}\mathcal{D}\left( \mathbf{L}_{k}^{T} \hat{\vec{\xi}} \, \right) \hat{\rho%
}\left( t\right)  \label{gaussian}
\end{equation}%
where $\mathbf{H}$ is a matrix form of the Hamiltonian, and the second term
describes the effects of dissipation. We have three dissipative baths (one
for each cavity, and the mechanical bath), and each can either absorb or
emit energy, hence the index $k$ runs from $1$ to $6$. For each term, $%
\gamma _{k}$ describes the rate of either energy absorption or emission by
one of the baths, and $\mathbf{L}_{k}^{T} \hat{\vec{\xi}}$\textbf{\ } is the
corresponding \textquotedblleft jump" operator (which is linear in the
system quadrature operators). $\mathcal{D}$ denotes a standard Linblad
superoperator: % OLD YINGDAN TEXT
%where $\mathbf{H}$ is a matrix form of the Hamiltonian, $\gamma _{k}$ is the
%damping rate of $k$-th dissipation channel, $1\leq k\leq 2N$, $\mathbf{L}%
%_{k}^{T}\mathbf{\hat{\xi}}$\textbf{\ }represents the jump operators and they
%are linear in terms of quadrature operators
\begin{equation}
\mathcal{D}\left( \mathbf{L}_{k}^{T} \hat{\vec{\xi}} \, \right) \hat{\rho}%
\left( t\right) \equiv 2 \left( \mathbf{L}_{k}^{T} \hat{\vec{\xi}} \,
\right) \hat{\rho}\left( t\right) \left( \mathbf{L}_{k}^{T} \hat{\vec{\xi}}
\, \right)^\dag -\left\{ \left( \mathbf{L}_{k}^{T} \hat{\vec{\xi}} \,
\right)^\dag \left( \mathbf{L}_{k}^{T} \hat{\vec{\xi}} \, \right) , \hat{\rho%
}\left( t\right) \right\}
\end{equation}%
Using Eq.~(\ref{gaussian}), one can get the time evolution of the average
values $\left\langle \mbox{\boldmath$\xi$}\right\rangle $ and the covariance
matrix $\mathbf{V}$:

\begin{equation}
\frac{d\left\langle \hat{\vec{\xi}}\left( t\right) \right\rangle }{dt}\equiv
\mathbf{Q}\left\langle \hat{\vec{\xi}}\left( t\right) \right\rangle ,\text{
\ }\frac{d\mathbf{V}}{dt}=\mathbf{QV}+\mathbf{VQ}^{T}+\mathbf{N}
\label{gaueq}
\end{equation}%
with
\begin{equation}
\mathbf{Q}=2\mathbf{\mbox{\boldmath${\sigma}$}H}+2\mathbf{%
\mbox{\boldmath${\sigma}$}}\left( \text{Im }\mathbf{\mbox{\boldmath${%
\Gamma}$}}\right) \text{, }\mathbf{N}=2\mathbf{\mbox{\boldmath${\sigma}$}}%
\left( \text{Re}\mathbf{\mbox{\boldmath${\Gamma}$}}\right) \mathbf{%
\mbox{\boldmath${\sigma}$}}^{T}  \label{qn}
\end{equation}%
and the elements of $\mathbf{\sigma }$ and $\mathbf{\Gamma }\ $are defined
as
\begin{equation}
\sigma _{ij}=-i\left[ \hat{\xi}_{i},\hat{\xi}_{j}\right] \text{, \ }\Gamma
_{mn}=\sum_{k}\frac{\gamma _{k}}{2}L_{k,m}^{\ast }L_{k,n}
\end{equation}%
Solving Eq. (\ref{gaueq}), the solutions are
\begin{eqnarray}
\left\langle \hat{\vec{\xi}}\left( t\right) \right\rangle &=&e^{\mathbf{Q}%
t}\left\langle \hat{\vec{\xi}}\left( 0\right) \right\rangle  \notag \\
\mathbf{V}\left( t\right) &=&e^{\mathbf{Q}t}\,\mathbf{V}\left( 0\right) e^{%
\mathbf{Q}^{T}t}+\int_{0}^{t}d\tau \,e^{\mathbf{Q}\left( t-\tau \right) }%
\mathbf{N}e^{\mathbf{Q}^{T}\left( t-\tau \right) }  \label{te}
\end{eqnarray}%
Using these results, one can calculate $\bar{n}_{\mathrm{h}}$ and $\lambda $
in Eq. (\ref{nlam}) and the fidelity of state transfer. In the following, we
show the result for double swap scheme, hybrid scheme and itinerant state
transfer. As for the adiabatic passage scheme, high fidelity transfer
requires a trade-off between the adiabaticity and the fast operation. The
optimal evolution is thus not purely in the adiabatic limit; as a result,
simple analytic expressions are difficult to obtain.

\subsection{Double swap scheme}

\label{app:doubleswap}

For the double swap scheme, the interaction is switched on sequentially. At
each swap process, the relevant evolution and noise matrix can be derived
from Eq. (\ref{qn}) to be
\begin{equation}
\mathbf{Q}_{i}=G_{i}\mathbf{S}-\frac{\kappa _{i}+\gamma }{4}\mathbf{1}+\frac{%
\kappa _{i}-\gamma }{4}\mathbf{Y}\text{, }\mathbf{N}=\frac{\bar{\gamma}+\bar{%
\kappa}_{i}}{8}\mathbf{1}+\frac{\bar{\gamma}-\bar{\kappa}_{i}}{8}\mathbf{Y,}
\label{qds}
\end{equation}%
with $\bar{\gamma}=\gamma \left( 2N_{m}+1\right) $, $\bar{\kappa}_{i}=\kappa
_{i}\left( 2N_{i}+1\right) $ and%
\begin{equation}
\mathbf{S}=\left(
\begin{array}{cccc}
0 & 0 & 0 & 1 \\
0 & 0 & -1 & 0 \\
0 & 1 & 0 & 0 \\
-1 & 0 & 0 & 0%
\end{array}%
\right) ,\mathbf{Y=}\left(
\begin{array}{cccc}
1 & 0 & 0 & 0 \\
0 & 1 & 0 & 0 \\
0 & 0 & -1 & 0 \\
0 & 0 & 0 & -1%
\end{array}%
\right) .  \label{sy}
\end{equation}%
Here the subspace is the direct product of two systems under swap, i.e, in
the first swap, the subspace is the direct product of the cavity 1 and the
mechanical resonator; while the subspace of the second swap is the direct
product of the mechanical resonator and the cavity 2.

The initial state in cavity 1 to be transferred is assumed to be a squeezed
state%
\begin{equation}
\left\vert \alpha ,r\right\rangle =\hat{D}\left[ \alpha \right] \hat{S}\left[
r\right] \left\vert 0\right\rangle
\end{equation}%
with%
\begin{equation}
\hat{S}\left[ r\right] \equiv \exp \left[ \frac{r}{2}\hat{d}_{1}^{2}-\frac{r%
}{2}\hat{d}_{1}^{\dag 2}\right] ,
\end{equation}%
where $r$ is taken to be real, and
\begin{equation}
\hat{D}\left[ \alpha \right] =\exp \left[ \alpha \hat{d}_{1}^{\dag }-\alpha
^{\ast }\hat{d}_{1}\right] .
\end{equation}%
Then the initial mean value and covariance matrix are
\begin{equation}
\left\langle \hat{\vec{\xi}}_{\mathrm{i}}\right\rangle =-\left(
\begin{array}{c}
\text{Re }\alpha \\
\text{Im }\alpha%
\end{array}%
\right) ,\text{ \ }\mathbf{V}_{\mathrm{i}}=\frac{1}{4}\left(
\begin{array}{cc}
e^{-2r} & 0 \\
0 & e^{2r}%
\end{array}%
\right) .  \label{xi}
\end{equation}%
The initial state of the mechanical resonator is assumed to be pre-cooled
into the ground state and the cavity 2 is also assumed to be in ground state
initially. The swap time $t_{i\text{s}}=\pi /\left( 2G_{i}\right) $. Then
the time evolution at $t_{\text{s}}\equiv \sum_{i}t_{i\text{s}}$ ($i=1,2$)
is given by Eq. (\ref{te}) as:
\begin{equation}
\left\langle \hat{\vec{\xi}}_{\mathrm{f}}\left( t_{\text{s}}\right)
\right\rangle =-\left(
\begin{array}{c}
\left( \text{Re }\alpha \right) e^{-\sum_{i}\varkappa _{i}t_{i\text{s}}} \\
\left( \text{Im }\alpha \right) e^{-\sum_{i}\varkappa _{i}t_{i\text{s}}}%
\end{array}%
\right)  \label{xf}
\end{equation}%
\begin{equation}
\mathbf{V}_{\mathrm{f}}\left( t_{\text{s}}\right) =\frac{e^{-2\sum_{i}%
\varkappa _{i}t_{i\text{s}}}}{4}\left(
\begin{array}{cc}
e^{-2r} & 0 \\
0 & e^{2r}%
\end{array}%
\right) \mathbf{+}\frac{B+e^{-2\varkappa _{2}t_{2\text{s}}}\left( \nu
_{2}^{2}+e^{-2\varkappa _{1}t_{1\text{s}}}\nu _{1}^{2}+A\right) }{4}\mathbf{1%
}  \label{vf}
\end{equation}%
with $\varkappa _{i}=\left( \kappa _{i}+\gamma \right) /4$, $\nu _{i}=\left(
\kappa _{i}-\gamma \right) /\left( 4G_{i}\right) $, $A=\bar{\kappa}_{1}\mu
_{1}+\bar{\gamma}\alpha _{1}$, $B=\bar{\kappa}_{2}\beta _{2}+\bar{\gamma}\mu
_{2}$ and
\begin{eqnarray}
\alpha _{i} &\approx &\int_{0}^{t_{i\text{s}}}d\tau \left( \cos G_{i}\tau
+\nu _{i}\sin G_{i}\tau \right) ^{2}e^{-\varkappa _{i}\tau }  \notag \\
\beta _{i} &\approx &\int_{0}^{t_{i\text{s}}}d\tau \left( \cos G_{i}\tau
-\nu _{i}\sin G_{i}\tau \right) ^{2}e^{-\varkappa _{i}\tau }  \notag \\
\mu _{i} &\approx &\int_{0}^{t_{i\text{s}}}d\tau \sin ^{2}G_{i}\tau
e^{-\kappa _{i}\tau }
\end{eqnarray}%
where we have assumed the strong coupling limit and hence approximate $\sqrt{G^{2}-\left( \kappa -\gamma \right) ^{2}/4}\approx G$. Note that if
pre-cooling is not performed, in Eq. (\ref{vf}) $\nu _{1}^{2}\rightarrow \nu
_{1}^{2}\left( 2N_{M,0}+1\right) $, with $N_{M,0}$ the initial phonon number
of the mechanics.

Plugging Eqs.~(\ref{xi}), (\ref{xf}) and (\ref{vf})) into Eqs.~(\ref%
{eq:FullFidelity}), (\ref{nb}) and (\ref{nlam}), we obtain the full fidelity
for transferring a coherent state. In the relevant strong-coupling limit
where $\kappa _{i},\gamma \ll G$, this results in Eqs.~(\ref{ds}) and (\ref{nwoc1}) in the main text.

The calculation of the fidelity is only slightly more complicated in the
case of a displaced, squeezed state, where now the parameter $r>0$
(c.f.~Eq.~(\ref{xi})). Keeping the phase of the displacement (i.e.~$\alpha
=|\alpha |e^{i\phi }$), the two parameters determining $F$ become (keep
until the leading order of $\left( \kappa _{i}+\gamma \right) /G$, also keep
the higher-order terms of $\left( \kappa _{i}N_{i}+\gamma N_{m}\right) /G$
as $N_{m}$ can be large):%
\begin{eqnarray}
\bar{n}_{\mathrm{h}}[r] &\approx &\sqrt{1+2\bar{n}_{h}\left[ 0\right] \cosh
2r+\bar{n}_{h}^{2}\left[ 0\right] +2\left( \cosh 2r-1\right) \sum_{i}\varkappa _{i}t_{i\text{s}}}-1\text{,}
\label{eq:FullSqueezednh} \\
\text{ }\lambda \lbrack r] &\approx &\lambda \lbrack 0]\sqrt{\frac{\cos
^{2}\phi }{e^{-2r}+\bar{n}_{\mathrm{h}}[0]}+\frac{\sin ^{2}\phi }{e^{2r}+
\bar{n}_{\mathrm{h}}[0]}}
\end{eqnarray}%
where $\bar{n}_{\mathrm{h}}[0]$, $\lambda \lbrack 0]$ are heating and
amplitude-decay parameters for a coherent state (c.f.~Eq.~(\ref{ds})).
Consider a squeezed ground state (i.e.~$\alpha =0$) in the strong squeezing
limit $r\rightarrow \infty $. In this case, there is no amplitude-decay
contribution to the fidelity $\lambda \lbrack r]=0$, and
\begin{equation}
\bar{n}_{\mathrm{h}}[r]\rightarrow e^{r}\sqrt{\bar{n}_{h}+ \sum_{i}\varkappa _{i}t_{i\text{s}}}  \label{nsqu}
\end{equation}%
We see that the effects of mechanical heating are exponentially enhanced.
Thus, for highly squeezed states (or other states with fine structures in
phase space), a state transfer scheme which attempts to suppress these
effects by using the mechanically-dark mode will be much better than the
double-swap scheme.

\subsection{Hybrid scheme}

\label{app:hybrid}

Both coupling $G_{1}$ and $G_{2}$ are switched on at the same time in hybrid
scheme. The relevant evolution and noise matrix can be derived from Eq.~(\ref%
{qn}) to be%
\begin{eqnarray}
\mathbf{Q} &\approx &G\mathbf{S-}\left( \frac{\kappa +\gamma }{4}\mathbf{1}+%
\frac{\kappa -\gamma }{4}\mathbf{Y}\right) ,  \notag \\
\mathbf{N} &\mathbf{=}&\frac{\bar{\kappa}}{8}\left( \mathbf{1+Y}\right) +%
\frac{\bar{\gamma}}{8}\left( \mathbf{1-Y}\right)
\end{eqnarray}%
where we have taken $G_{1}=G_{2}=G$, $\kappa _{1}=\kappa _{2}=\kappa $, $%
\bar{\gamma}=\gamma \left( 2N_{m}+1\right) $, $\bar{\kappa}_{i}\approx \bar{%
\kappa}$, $\mathbf{1}$ is a $6\times 6$ unity matrix and the definitions of $%
\mathbf{S}$ and $\mathbf{Y}$ are similar as Eq. (\ref{sy}) except that the
dimension is increased to $6\times 6$:%
\begin{equation}
\mathbf{S}=\left(
\begin{array}{cccccc}
0 & 0 & 0 & 1 & 0 & 0 \\
0 & 0 & -1 & 0 & 0 & 0 \\
0 & 1 & 0 & 0 & 0 & 1 \\
-1 & 0 & 0 & 0 & -1 & 0 \\
0 & 0 & 0 & 1 & 0 & 0 \\
0 & 0 & -1 & 0 & 0 & 0%
\end{array}%
\right) ,\mathbf{Y=}\left(
\begin{array}{cccccc}
1 &  &  &  &  &  \\
& 1 &  &  &  &  \\
&  & -1 &  &  &  \\
&  &  & -1 &  &  \\
&  &  &  & 1 &  \\
&  &  &  &  & 1%
\end{array}%
\right)
\end{equation}%
Then starting from a squeezed state as described by Eq.~(\ref{xi}), the time
evolution at $t_{\text{hs}}\equiv \pi /v$ (with $\nu =\sqrt{2G^{2}-\left(
\kappa -\gamma \right) ^{2}/16}$ ) is given by Eq.~(\ref{te}) as:
\begin{equation}
\left\langle \hat{\vec{\xi}}_{\mathrm{f}}\left( t_{\text{hs}}\right)
\right\rangle =-\frac{e^{-\frac{\kappa +\gamma }{4}t_{\text{hs}}}}{2}\left(
1+e^{-\frac{\kappa -\gamma }{4}t_{\text{hs}}}\right) \left(
\begin{array}{c}
\text{Re }\alpha  \\
\text{Im }\alpha
\end{array}%
\right)
\end{equation}%
\begin{eqnarray}
\mathbf{V}_{\mathrm{f}}\left( t_{\text{hs}}\right)  &=&\frac{1}{4}\left(
\frac{e^{-\frac{\kappa +\gamma }{2}t_{\text{hs}}}}{4}\left( 1-e^{-\frac{%
\kappa -\gamma }{4}t_{\text{hs}}}\right) ^{2}+\delta \frac{G^{2}}{\nu ^{2}}%
\bar{\gamma}+\left( \frac{\delta \kappa ^{2}}{16\nu ^{2}}-\frac{\xi \kappa }{%
2\nu }+\chi +\eta \right) \frac{\bar{\kappa}}{2}\right) \mathbf{1}  \notag \\
&&+\frac{e^{-\frac{\kappa +\gamma }{2}t_{\text{hs}}}}{16}\left( 1+e^{-\frac{%
\kappa -\gamma }{4}t_{\text{hs}}}\right) ^{2}\left(
\begin{array}{cc}
e^{-2r} & 0 \\
0 & e^{2r}%
\end{array}%
\right)
\end{eqnarray}%
with
\begin{eqnarray}
\text{ }\delta  &=&\int_{0}^{t_{\text{hs}}}d\tau e^{-\frac{\kappa +\gamma }{2%
}\tau }\sin ^{2}\nu \tau ,\text{ \ \ \ \ \ \ \ \ }\eta =\int_{0}^{t_{\text{hs%
}}}d\tau e^{-\kappa \tau },  \notag \\
\xi  &=&\int_{0}^{t_{\text{hs}}}d\tau e^{-\frac{\kappa +\gamma }{2}\tau
}\sin \nu \tau \cos \nu \tau ,\text{ \ }\chi =\int_{0}^{t_{\text{hs}}}d\tau
e^{-\frac{\kappa +\gamma }{2}\tau }\cos ^{2}\nu \tau \text{.}
\end{eqnarray}%
In the strong coupling limit $\kappa \ll G$, and considering a coherent
state $r=0$, the heating and damping can be simplified to Eq.~(\ref{hb}) in
the main text.

\section{Transfer fidelity of itinerant Gaussian state}

\label{app:itinerant}

In order to calculate the transfer fidelity for itinerant Gaussian state, we
first define the input state to be generated by the mode creation operator $\hat{D}_\text{in}$ ($\hat{D}_\text{in}$ can be $\hat{D}_{1,\text{in}}$, $\hat{D}_{2,\text{in}}$)
which represents a time mode which spreads over a frequency  interval $\Delta \omega $ around frequency $\omega _{M}$ and localized in time around $t=0$
\begin{equation}
\hat{D}_{\text{in}}=\left( 2\pi \right) ^{-1/2}\int d\omega f\left[ \omega %
\right] \hat{d}_{\text{in}}\left[ \omega \right]   \label{din1}
\end{equation}%
with $\hat{d}_{\text{in}}\left[ \omega \right] $ is the Fourier transform
of\ the input operator which satisfies $\left[ \hat{d}_{\text{in}}\left[
\omega _{1}\right] ,\hat{d}^\dag_{\text{in}}\left[ \omega _{2}\right] \right] $ $%
=2\pi \delta \left( \omega _{1}-\omega _{2}\right) $, and $f(\omega)$ is normalized function which is localized in both frequency and time.
%\begin{equation}
%f\left[ \omega \right] =\left( \frac{1}{2\pi \Delta \omega ^{2}}\right)
%^{1/4}\exp \left( -\frac{\left( \omega -\omega _{M}\right) ^{2}}{4\Delta
%\omega ^{2}}\right) .
%\end{equation}

We define a smoothed operator $\hat{d}_{\text{in}}^{\prime }\left[ \omega
_{j}\right] $ with discrete frequencies $\omega _{j}=j\cdot \delta _{\omega }
$ ($j\in \mathbb{Z}$)~\cite{ClerkRMP}%
\begin{equation}
\hat{d}_{\text{in}}^{\prime }\left[ \omega _{j}\right] =\frac{1}{\sqrt{2\pi
\delta _{\omega }}}\int_{\omega _{j}}^{\omega _{j+1}}d\omega \hat{d}_{\text{%
in}}\left[ \omega \right] .
\end{equation}%
Here $\delta _{\omega }$ is a small frequency interval on which $f\left[
\omega \right] $ is approximately constant, and $\hat{d}_{\text{in}}^{\prime
}\left[ \omega _{j}\right] $ satisfies the normal bosonic commutation
relation
\begin{equation}
\left[ \hat{d}_{\text{in}}^{\prime }\left[ \omega _{j}\right] ,\hat{d}_{%
\text{in}}^{\prime \dag }\left[ \omega _{k}\right] \right] =\delta _{jk}.
\end{equation}%
The integral of Eq. (\ref{din1}) is then changed into the summation
\begin{equation}
\hat{D}_{\text{in}}\equiv \sqrt{2\pi \delta _{\omega }}\sum_{j}f\left[
\omega _{j}\right] \hat{d}_{\text{in}}^{\prime }\left[ \omega _{j}\right]. \label{dsum}
\end{equation}

The output operator can be defined as
\begin{equation}
\hat{D}_{\text{out}}=\left( 2\pi \right) ^{-1/2}\int d\omega f\left[ \omega %
\right] e^{-i(\omega -\omega _{M})\tau }\hat{d}_{\text{out}}\left[ \omega %
\right]   \label{dout1}
\end{equation}%
with $\tau $ is the time delay of the wave packet and a discretized summation can also be defined in a similar fashion as Eq.~(\ref{dsum}).

Suppose the initial state is prepared in a coherent state
\begin{equation}
\hat{\rho}=\left\vert \alpha \right\rangle _{1}\left\langle \alpha
\right\vert \otimes \left\vert 0\right\rangle _{2}\left\langle 0\right\vert
\otimes \hat{\rho}_{a,\text{th}}  \label{r0}
\end{equation}%
with%
\begin{equation}
\left\vert \alpha \right\rangle _{1}=\exp \left[ \alpha \hat{D}_{1,\text{in}%
}^{\dag }-\alpha ^{\ast }\hat{D}_{1,\text{in}}\right] \left\vert
0\right\rangle .
\end{equation}%
The output state in Heisenberg picture is the same as Eq.~(\ref{r0}), but
the operators are changed from ``in" to ``out" according to the scattering
matrix
\begin{equation}
\hat{d}_{2,\text{out}}\left[ \omega \right] =s_{21}\left[ \omega \right]
\hat{d}_{1,\text{in}}\left[ \omega \right] +s_{22}\left[ \omega \right] \hat{%
d}_{2,\text{in}}\left[ \omega \right] +s_{23}\left[ \omega \right] \hat{a}_{%
\text{in}}\left[ \omega \right]
\end{equation}

Using Eq.~(\ref{din1}), (\ref{dout1}), and taking the limit $\delta _{\omega
}\rightarrow 0$ and change the summation back to integral, we obtain the
average value of quadrature in state Eq.~(\ref{r0}) as:
\begin{equation}
\left\langle \hat{\vec{\xi}}_{1,\text{in}}\right\rangle =\left(
\begin{array}{c}
\text{Re }\alpha  \\
\text{Im }\alpha
\end{array}%
\right) ,\left\langle \hat{\vec{\xi}}_{2,\text{out}}\right\rangle =\left(
\begin{array}{c}
\text{Re }\alpha \zeta _{1} \\
\text{Im }\alpha \zeta _{1}%
\end{array}%
\right) ,
\end{equation}%
where
\begin{equation}
\zeta _{1}=\left( 2\pi \int d\omega e^{-i\omega \tau }s_{21}\left[ \omega %
\right] \left\vert f\left[ \omega \right] \right\vert ^{2}\right) _{\omega
_{M}=0}.
\end{equation}%
And the covariance matrix are
\begin{equation}
\mathbf{V}_{1,\text{in}}=\frac{1}{4}\left(
\begin{array}{cc}
1 & 0 \\
0 & 1%
\end{array}%
\right) ,\text{ }\mathbf{V}_{2,\text{out}}=\frac{1}{4}\left( 2\bar{n}_{\mathrm{h}}+1\right)
\left(
\begin{array}{cc}
1 & 0 \\
0 & 1%
\end{array}%
\right)
\end{equation}%
Using Eq.~(\ref{nb}) and (\ref{nlam}), we obtain the heating and amplitude damping parameters as shown Eq.~(\ref{nl}) in the main text.

\section{Transfer fidelity of itinerant non-Gaussian state}

\label{app:nongaussian}

A Fock state input of cavity 1 is defined as
\begin{equation}
\hat{\rho}_{1,\text{in}}=\frac{1}{n!}\left( \hat{D}_{1,\text{in}}^{\dag
}\right) ^{n}\left\vert 0\right\rangle _{1}\left\langle 0\right\vert \left(
\hat{D}_{1,\text{in}}\right) ^{n}
\end{equation}%
where $\left\vert 0\right\rangle _{1}$ is the vacuum state of the input mode
and we only consider the input pulse with zero bandwidth for simplicity. We
also assume the state of the cavity 2 is vacuum $\hat{\rho}%
_{2,in}=\left\vert 0\right\rangle _{2}\left\langle 0\right\vert ,$ while the
mechanical resonator is in a thermal state
\begin{equation}
\hat{\rho}_{a,\text{in}}=\sum_{q}p_{q}\frac{1}{n!}\left( \hat{a}_{\text{in}%
}^{\dag }\right) ^{q}\left\vert 0\right\rangle _{a}\left\langle 0\right\vert
\left( \hat{a}_{\text{in}}\right) ^{q}
\end{equation}%
with $p_{q}=\left( 1-e^{-\beta \hbar \omega }\right) e^{-\beta \hbar \omega
q}$. In terms of output operator, the output state is
\begin{equation}
\hat{\rho}_{\text{out}}=\hat{\rho}_{\text{in}}=\sum_{q}p_{q}\left\vert \psi
_{q}\right\rangle \left\langle \psi _{q}\right\vert
\end{equation}%
with%
\begin{equation}
\left\vert \psi _{q}\right\rangle =\frac{1}{\sqrt{n!q!}}\left( u_{11}\hat{D}%
_{1,\text{out}}^{\dag }+u_{12}\hat{D}_{2,\text{out}}^{\dag }+u_{13}\hat{a}_{%
\text{out}}^{\dag }\right) ^{n}\left( u_{31}\hat{D}_{1,\text{out}}^{\dag
}+u_{32}\hat{D}_{2,\text{out}}^{\dag }+u_{33}\hat{a}_{\text{out}}^{\dag
}\right) ^{q}\left\vert 0\right\rangle .  \label{pq}
\end{equation}%
and $\mathbf{u}=\mathbf{s}^{-1}$ is the inverse of the scattering matrix.
Hence the fidelity can be written as%
\begin{equation}
F=\sum_{q}p_{q}\left( \sum_{n_{1},q_{a}}\left\vert \left\langle \psi
_{q}|n_{1},n_{2},q_{a}\right\rangle \right\vert ^{2}\right)
\end{equation}

According to Eq.~(\ref{pq}), both $u_{11}$ ($o\left( 1/C\right) $) and $%
u_{31}$ ($o\left( 1/\sqrt{C}\right) $) are small, i.e. all the processes to
generate photon (reflecting) in cavity 1 are perturbations. One can
therefore expand the state in terms of the photon numbers from the output of
cavity 1. Hence\ if we expand the fidelity as the summation of the
probability of having $r$ ($r=0,1,2...$) photon in the cavity 1 output, the
series will converge quickly for large cooperativity.

After some lengthy but simple algebraic calculations, we get the probability
to have $0$ photon in cavity 1 output%
\begin{eqnarray}
F^{\left( 0\right) } &=&\sum_{q}p_{q}\left\vert u_{12}\right\vert
^{2n}\left\vert u_{33}\right\vert ^{2q}\left\vert {_{2}F_{1}} \left( -n,-q;1;
\frac{u_{13}u_{32}}{u_{12}u_{33}}\right) \right\vert ^{2} \\
F^{\left( 1\right) } &=&\sum_{q}p_{q}q\left\vert u_{33}\right\vert
^{2q-2}\left\vert u_{12}\right\vert ^{2n}\left\vert
\begin{array}{c}
n\frac{u_{11}u_{32}}{u_{12}} \cdot \ {_{2}F_{1}} \left( 1-n,1-q;2;\frac{%
u_{13}u_{32}}{u_{12}u_{33}}\right) \\
+u_{31}\cdot \ {_{2}F_{1}} \left( -n,1-q;1;\frac{u_{13}u_{32}}{u_{12}u_{33}}
\right)%
\end{array}%
\right\vert
\end{eqnarray}%
where $F_{\left( 2,1\right) }$ is hypergeometric function. $F^{\left(
0\right) }$ is a lower bound of the exact fidelity, which physically
corresponds to the case that the two cavities swap their states, so this
fidelity represents a "swapping fidelity". Expansion of $F^{\left( 0\right)
} $ in terms of $\varepsilon $ gives a contribution in the fidelity which is
precise up to $o\left( 1/C^{2}\right) $,
\begin{equation}
F^{\left( 0\right) }\approx \sum_{q}p_{q}\left( 1-\left( \frac{n}{2}%
+q+nq\right) \varepsilon ^{2}\right) ^{2}  \label{f0a}
\end{equation}%
The probability to have $1$ photon in cavity 1 output is
\begin{equation}
F^{\left( 1\right) }=\sum_{q}p_{q}q\left\vert u_{33}\right\vert
^{2q-2}\left\vert u_{12}\right\vert ^{2n}\left\vert
\begin{array}{c}
n\frac{u_{11}u_{32}}{u_{12}}\cdot \ {_{2}F_{1}}\left( 1-n,1-q;2;\frac{%
u_{13}u_{32}}{u_{12}u_{33}}\right) \\
+u_{31}\cdot \ {_{2}F_{1}}\left( -n,1-q;1;\frac{u_{13}u_{32}}{u_{12}u_{33}}%
\right)%
\end{array}%
\right\vert ^{2}.
\end{equation}%
$F^{\left( 0\right) }+F^{\left( 1\right) }$ is a better lower bound of the
actual fidelity. Physically it represents the fidelity of state transfer if
the photon reflected from cavity 1 is measured to be 0 or 1.

Generally, to have $r$ photons in the output of cavity 1, the corresponding
probability is
\begin{eqnarray}
F^{\left( r\right) } &=&\sum_{q}p_{q}\left\vert u_{12}\right\vert
^{2n}\left\vert u_{33}\right\vert ^{2q}\left(
\begin{array}{c}
q \\
r%
\end{array}%
\right) \cdot  \notag \\
&&\left\vert \sum_{j=0}^{r}\left(
\begin{array}{c}
n \\
j%
\end{array}%
\right) \left(
\begin{array}{c}
r \\
j%
\end{array}
\right) \left( \frac{u_{31}}{u_{33}}\right) ^{r-j}\left( \frac{u_{11}u_{32}}{%
u_{12}u_{31}}\right) ^{j}{_{2}F_{1}}\left( j-n,r-q;1+j;\frac{u_{13}u_{32}}{%
u_{12}u_{33}}\right) \right\vert ^{2}
\end{eqnarray}
The total fidelity is a summation of the all the probabilities $%
F=\sum_{r}F^{\left( r\right) }$. In terms of scattering matrix, we get Eq.~(%
\ref{fock}) in the main text.

Furthermore, if the input state is a more general state $\hat{\rho}
_{in,1}=\sum_{mn}c_{n}^{\ast }c_{m}\left\vert m\right\rangle \left\langle
n\right\vert $, following a similar analysis, we can get Eq.~(\ref{fs}) in
the main text.

\section*{References}

%\bibliographystyle{iopart-num}
%\bibliography{statetransfer}
%\begin{thebibliography}{10}
\providecommand{\newblock}{}

\end{document}